\documentclass{emulateapj}
\usepackage{amsmath}
\usepackage{array,multirow,natbib,color,tablefootnote}
\usepackage[flushleft]{threeparttable}
\usepackage{float}
\usepackage{graphicx}
\usepackage{subfigure}
\usepackage{setspace}
\usepackage{epstopdf,textcomp}

\newcommand{\ts}{\textsuperscript}
\newcommand{\bjdtdb}{\ensuremath{\rm {BJD_{TDB}}}}

\newcommand{\ms}{m s$^{-1}$}

\bibpunct[ ]{(}{)}{;}{}{}{,}
\shorttitle{WASP-14b Phase Curves}
\shortauthors{Wong et al.}

\begin{document}
\title{3.6 and 4.5~$\mu$\MakeLowercase{m} Phase Curves of the Highly-Irradiated Eccentric Hot Jupiter WASP-14\MakeLowercase{b}}

\author{Ian Wong,\altaffilmark{1}\altaffilmark{*} Heather A. Knutson,\altaffilmark{1} Nikole K. Lewis,\altaffilmark{2} Tiffany Kataria,\altaffilmark{3} Adam Burrows,\altaffilmark{4} Jonathan J. Fortney,\altaffilmark{5} Joel Schwartz,\altaffilmark{6} Eric Agol,\altaffilmark{7} Nicolas B. Cowan,\altaffilmark{8} Drake Deming,\altaffilmark{9}  Jean-Michel D{\' e}sert,\altaffilmark{10}  Benjamin J. Fulton,\altaffilmark{11,12} Andrew W. Howard,\altaffilmark{11}  Jonathan Langton,\altaffilmark{13} Gregory Laughlin,\altaffilmark{5}  Adam P. Showman,\altaffilmark{14} Kamen Todorov\altaffilmark{15}}

\altaffiltext{1}{Division of Geological and Planetary Sciences, California Institute of Technology, Pasadena, CA 91125, USA}
\altaffiltext{*}{iwong@caltech.edu}
\altaffiltext{2}{Space Telescope Science Institute, Baltimore, MD 21218, USA}
\altaffiltext{3}{Astrophysics Group, School of Physics, University of Exeter, Stocker Road, Exeter EX4 4QL, UK}
\altaffiltext{4}{Department of Astrophysical Sciences, Princeton University, Princeton, NJ 08544, USA}
\altaffiltext{5}{Department of Astronomy and Astrophysics, University of California at Santa Cruz, Santa Cruz, CA 95604, USA}
\altaffiltext{6}{Department of Physics \& Astronomy, Northwestern University, Evanston, IL 60208, USA}
\altaffiltext{7}{Department of Astronomy, University of Washington, Seattle, WA 98195, USA}
\altaffiltext{8}{Department of Physics and Astronomy, Amherst College, Amherst, MA 01002, USA}
\altaffiltext{9}{Department of Astronomy, University of Maryland, College Park, MD 20742, USA}
\altaffiltext{10}{Department of Astrophysical and Planetary Science, University of Colorado, Boulder, CO 80309, USA}
\altaffiltext{11}{Institute for Astronomy, University of Hawaii, Honolulu, HI 96822,USA}
\altaffiltext{12}{NSF Graduate Research Fellow}
\altaffiltext{13}{Department of Physics, Principia College, Elsah, IL 62028, USA}
\altaffiltext{14}{Lunar and Planetary Laboratory, University of Arizona, Tucson, AZ 85721, USA}
\altaffiltext{15}{Institute for Astronomy, ETH Z{\" u}rich, 8093 Z{\" u}rich, Switzerland}

\begin{abstract}
We present full-orbit phase curve observations of the eccentric ($e\sim 0.08$) transiting hot Jupiter WASP-14b obtained in the 3.6 and 4.5~$\mu$m bands using the \textit{Spitzer Space Telescope}. We use two different methods for removing the intrapixel sensitivity effect and compare their efficacy in decoupling the instrumental noise. Our measured secondary eclipse depths of $0.1882\%\pm 0.0048\%$ and $0.2247\%\pm 0.0086\%$ at 3.6 and 4.5~$\mu$m, respectively, are both consistent with a blackbody temperature of $2402\pm 35$~K. We place a $2\sigma$ upper limit on the nightside flux at 3.6~$\mu$m and find it to be $9\%\pm 1\%$ of the dayside flux, corresponding to a brightness temperature of 1079~K. At 4.5~$\mu$m, the minimum planet flux is $30\%\pm 5\%$ of the maximum  flux, corresponding to a brightness temperature of $1380\pm 65$~K. We compare our measured phase curves to the predictions of  one-dimensional radiative transfer and three-dimensional general circulation models. We find that WASP-14b's measured dayside emission is consistent with a model atmosphere with equilibrium chemistry and a moderate temperature inversion. These same models tend to over-predict the nightside emission at 3.6~$\mu$m, while under-predicting the nightside emission at 4.5~$\mu$m. We propose that this discrepancy might be explained by an enhanced global C/O ratio. In addition, we find that the phase curves of WASP-14b ($7.8~M_{\mathrm{Jup}}$) are consistent with a much lower albedo than those of other Jovian mass planets with thermal phase curve measurements, suggesting that it may be emitting detectable heat from the deep atmosphere or interior processes. \end{abstract}

\section{Introduction}\label{sec:intro}
Over the past two decades, observations of exoplanets have uncovered a stunning diversity of systems. Major improvements in the capabilities of space- and ground-based telescopes in recent years have led to the discovery and characterization of hundreds of new exoplanets, covering a broad range of orbital properties, interior and atmospheric compositions, and host star types \citep{han}. Meanwhile, these same technological advances have enabled the detailed study of the atmospheric properties of the brightest and largest planets through high-precision photometry and spectroscopy. Most of these targets belong to the class of gas giant planets known as hot Jupiters. The high levels of incident flux, slow rotation, and potentially large temperature gradients between hemispheres characteristic of these planets allow us to test atmospheric models in a new regime unlike any found in the Solar System. In addition, hot Jupiters are some of the most favorable targets for measuring elemental abundances, since most material is not in a condensed form at these high temperatures \citep[e.g.,][]{line}.

Atmospheric circulation models of hot Jupiters predict broad super-rotating equatorial jets that circulate energy between the day and night sides, with the precise effect of these winds on the day-to-night temperature contrast being strongly dependent on the particular orbital and atmospheric properties of the planet \citep[see][and references therein]{heng}. As a result of their short orbital periods, hot Jupiters have high atmospheric temperatures and emit relatively strongly at infrared wavelengths, allowing for direct measurement of their atmospheric brightness as a function of orbital phase.  These phase curves can then be converted into a longitudinal temperature profile \citep{cowanagol}. By comparing the measured phase curves to theoretical phase curves generated by atmospheric models, we can constrain fundamental properties of the atmosphere, such as the efficiency of heat transport from the day side to the night side, radiative time scales, wind speeds, and compositional gradients between the day and night sides. 

 To date, well-characterized phase curve observations have been published for eleven planets: $\upsilon$ And b \citep{harrington,crossfield}, HD 189733b \citep{knutson2007, knutson2009a, knutson2012}, HD 149026b \citep{knutson2009b}, HD 80606b \citep{laughlin}, HAT-P-7b \citep{borucki,welsh}, Kepler-7b \citep{demory}, CoRoT-1b \citep{snellen}, WASP-12b \citep{cowan2012}, WASP-18b \citep{maxted}, HAT-P-2b \citep{lewis}, and HD 209458b \citep{crossfield2012,zellem}. The majority of these observations were carried out using the \textit{Spitzer Space Telescope} while the rest were obtained at optical wavelengths by the \textit{CoRoT} and \textit{Kepler} missions. Recently, the first spectroscopic phase curve was obtained for the hot Jupiter WASP-43b using the \textit{Hubble Space Telescope} between 1.2 and 1.6 $\mu$m \citep{stevenson}.
 
In this paper, we present full-orbit phase curves of the hot Jupiter WASP-14b in the 3.6 and 4.5~$\mu$m bands obtained with the \textit{Spitzer Space Telescope}. Photometric and radial-velocity observations of WASP-14b indicate a mass of $M_{p}=7.3\pm 0.5~ M_{\mathrm{Jup}}$ and a radius of $R_{p}=1.28 \pm 0.08~R_{\mathrm{Jup}}$, corresponding to a density of $\rho=4.6~\mathrm{g/cm}^2$ \citep{joshi}. The planet lies on an eccentric orbit \citep[$e=0.0822\pm 0.003$;][]{knutson2014} around a young host star \citep[age $\sim 0.5-1.0$~Gyr, spectral type F5, $M_{*}=1.21\pm0.13~M_{\Sun}$, $R_{*}=1.31\pm0.07~R_{\Sun}$, 
$T_{*} = 6462 \pm 75$~K, and $\log{g} = 4.29\pm 0.04$;][]{joshi,torres}, with a period of 2.24 days and an orbital semi-major axis of  $a=0.036\pm 0.01$~AU. 

The equilibrium temperature of WASP-14b is relatively high \citep[$T_{\mathrm{eq}}= 1866$~K, assuming zero albedo and reemission from the entire surface;][]{joshi}, suggesting that the thermal evolution of WASP-14b may be significantly affected by Ohmic dissipation in a partially-ionized atmosphere (e.g., \citealp{batygin,batygin2013,perna}; but for debate see \citealp{rogers1,rogers2}). \citet{blecic} analyzed  secondary eclipse observations in the 3.6, 4.5, and 8.0$\mu$m \textit{Spitzer} bands and found no evidence for a thermal inversion in the dayside atmosphere, while concluding that the data are consistent with relatively poor energy redistribution from dayside to nightside. \citet{cowanagol2011} find that, based on the dayside fluxes, the most highly-irradiated planets have systematically less efficient day/night heat circulation. \citet{perez-becker} reached a similar conclusion from a comparison of the fractional day-night flux differences of planets for which phase curves have been obtained. This overall trend has been explained by hydrodynamical models \citep{perna2,perez-becker}.

Phase curve observations at more than one wavelength provide complementary information about the properties of the planet's atmosphere, as different wavelengths probe different pressure levels within the atmosphere. Multiband measurements of the planet's brightness can also be transformed into low-resolution dayside and nightside emission spectra, which can reveal differences in atmospheric composition. Only five systems have phase curve observations at more than one wavelength: HD 189733 \citep{knutson2012}, WASP-12 \citep{cowan2012}, WASP-18 \citep{maxted}, HAT-P-2 \citep{lewis}, and WASP-43 \citep{stevenson}. While single-wavelength phase curves can be reasonably well-matched by standard atmospheric circulation models, none of the multi-wavelength phase curves are satisfactorily reproduced by these same models. This suggests that our understanding of the physical and chemical processes that drive atmospheric circulation is still incomplete.

The paper is organized as follows: The observations and data reduction methodology are described in Section~\ref{sec:obs}. In Section~\ref{sec:analysis}, we discuss the phase curve model used in our analysis and present the best-fit parameters. We then use our results to obtain updated orbital parameters and discuss the implications of our phase curve fits for the planet's atmospheric dynamics in Section~\ref{sec:dis}.

\section{\textit{Spitzer} observations and photometry}\label{sec:obs}

\begin{figure}[b*]
\begin{center}
\includegraphics[width=9cm]{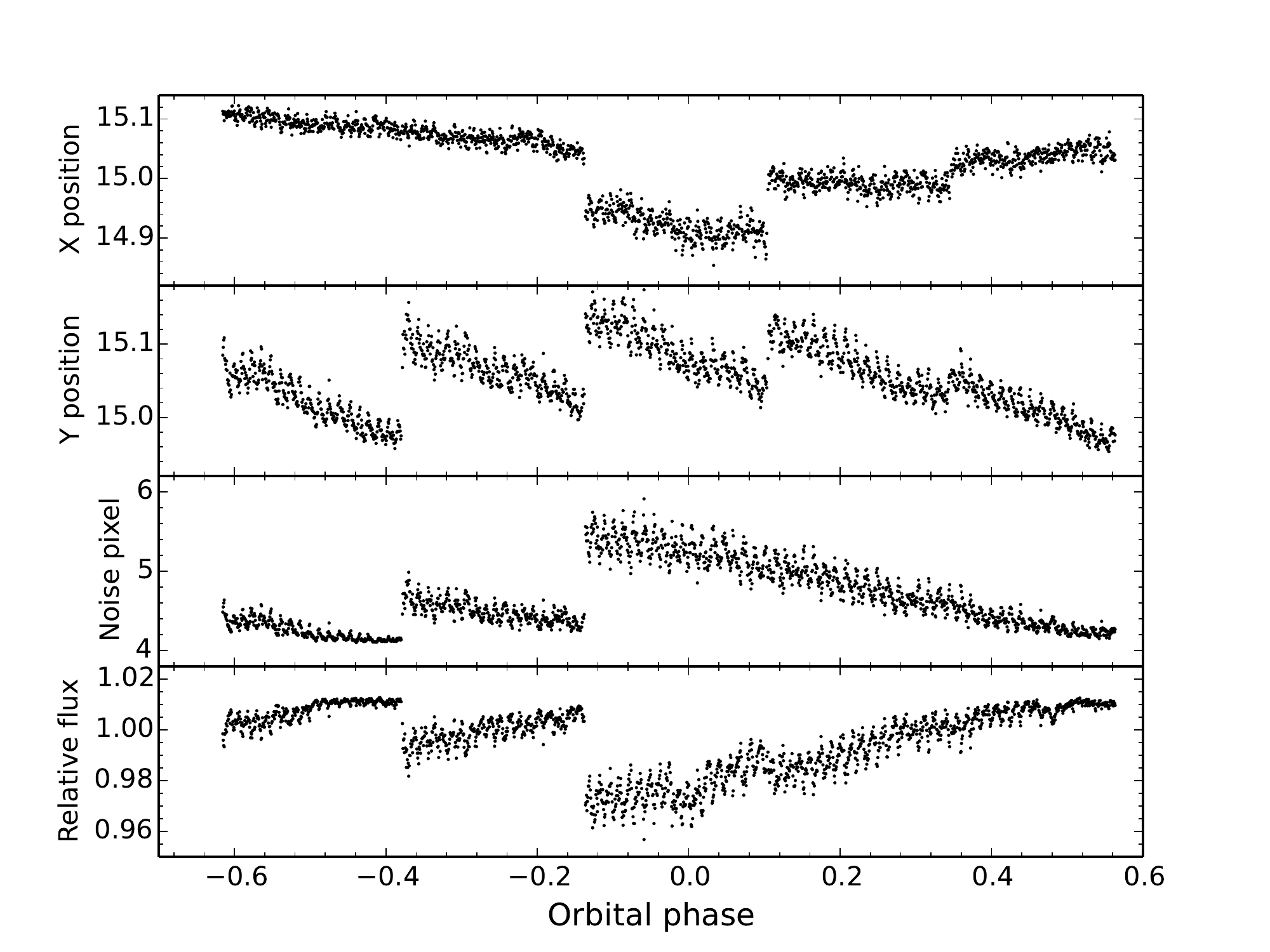}
\end{center}
\caption{Measured stellar $x$ centroids (top panel), $y$ centroids (upper middle panel), and noise pixel values (lower middle panel) as a function of orbital phase relative to transit for the 3.6~$\mu$m phase curve observation. The bottom panel shows the raw photometric series with hot pixels excised. The data are binned in 2-minute intervals.} \label{data1}
\end{figure}

\begin{figure}[t*]
\begin{center}
\includegraphics[width=9cm]{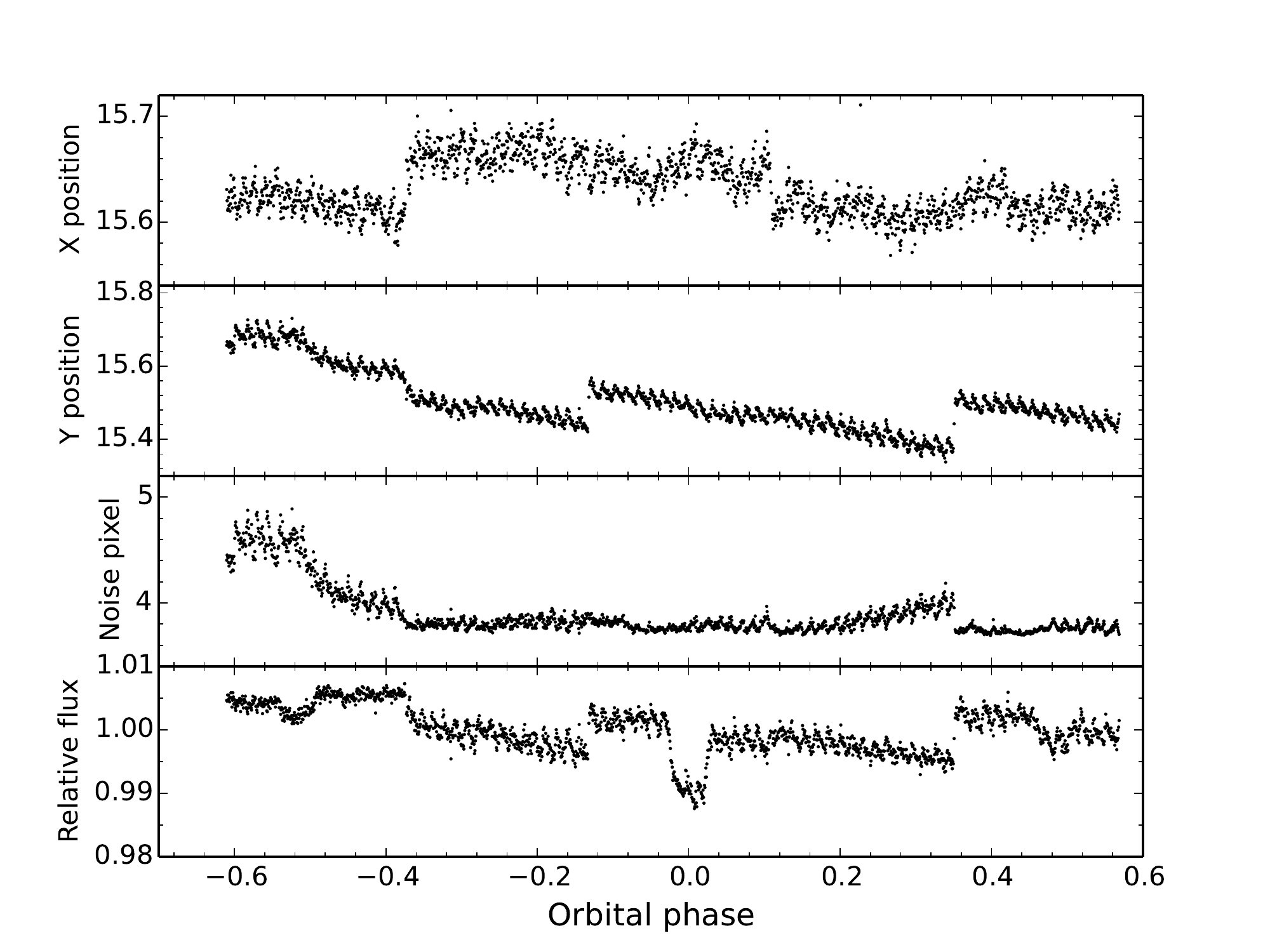}
\end{center}
\caption{Measured star centroids, noise pixel values, and raw photometric series for the 4.5~$\mu$m phase curve observations; see Figure~\ref{data1} for a complete description.} \label{data2}
\end{figure}

We observed two full orbits of WASP-14b in the 3.6 and 4.5~$\mu$m channels of the Infrared Array Camera \citep[IRAC;][]{fazio} on the \textit{Spitzer Space Telescope}. The observation periods were UT 2012 April 15$-$17 and UT 2012 April 24$-$26 for the 3.6 and 4.5~$\mu$m bandpasses, respectively. Both observations lasted approximately 64 hours and were carried out in subarray mode, which generated 32 $\times$ 32 pixel (39$''\times$39$''$) images with 2.0~s integration times, resulting in a total of 113,408 images in each bandpass. Due to long-term drift of the telescope pointing, the telescope was repositioned approximately every 12 hours in order to re-center the target, leading to four breaks in each phase curve observation with a combined duration of about 16 minutes. The telescope repositioning produced offsets of up to 0.2 pixels in the star's position on the array after each break (Figures~\ref{data1} and \ref{data2}).

We extract photometry using methods described in previous analyses of post-cryogenic \textit{Spitzer} data \citep[e.g.,][]{todorov,wong}. The raw data files are first dark-subtracted, flat-fielded, linearized, and flux-calibrated using version S19.1.0 of the IRAC pipeline. The exported data comprise a set of 1772 FITS files, each containing 64 images and a UTC-based Barycentric Julian Date (BJD$_{\mathrm {UTC}}$) time stamp designating the start of the first image. For each image, we calculate the BJD$_{\mathrm {UTC}}$ at mid-exposure by assuming uniform spacing and using the start and end times of each 64-image series as defined by the AINTBEG and ATIMEEND header keywords.

When estimating the sky background in each image, we avoid contamination from the wings of the star's point-spread function (PSF) by excluding pixels within a radius of 15 pixels from the center of the image, as well as the 13th-16th rows and the 14th and 15th columns, where the stellar PSF extends close to the edge of the array. In addition, we exclude the top (32nd) row of pixels, which have values that are consistently lower than those from other pixels in the array. We take the remaining set of pixels and iteratively trim values that lie more than $3\sigma$ from the median.  We then calculate the average sky background across the image by fitting a Gaussian function to the histogram of the remaining pixel values. After subtracting the sky background, any remaining transient ``hot pixels'' in each set of 64 images varying by more than $3\sigma$ from the median pixel value are replaced by the median pixel value. In both bandpasses, the average percentage of replaced pixels is less than $0.35\%$.

To determine the position of the star on the array in each image, we calculate the flux-weighted centroid for a circular region of radius $r_{0}$ pixels centered on the estimated position of the star \citep[see, for example,][]{knutson2008}. We then estimate the width of the star's PSF by computing the noise pixel parameter \citep{mighell}, which is defined in Section~2.2.2 of the \textit{Spitzer/IRAC} instrument handbook as
\begin{equation}\widetilde{\beta}=\frac{\left(\sum_{i}I_{i}\right)^2}{\sum_{i}I_{i}^2},\end{equation}
where $I_{i}$ is the intensity detected in the $i$th pixel. We define the parameter $r_{1}$ to be the radius of the circular aperture used to calculate $\widetilde{\beta}$.

We calculate the flux of the stellar target in each image using circular aperture photometry. We generate two sets of apertures: The first set uses a fixed aperture with radii ranging from 1.5 to 3.0 pixels in 0.1-pixel steps and from 3.0 to 5.0 pixels in 0.5-pixel steps. The second set utilizes a time-varying radius that is related to the square-root of the noise pixel parameter $\widetilde{\beta}$ by either a constant scaling factor or a constant shift \citep[see][for a full discussion of the noise-pixel-based aperture]{lewis}. The optimal choice of aperture photometry is determined by selecting data from 8000 images spanning the planetary transit with a total duration of 4.4~hours and calculating the photometric series for each choice of aperture. We then fit each photometric series with our transit light curve model (Section~\ref{subsec:eventmodel}), compute the RMS scatter in the resultant residuals binned in five-minute intervals, and choose the values of $r_{0}$ and $r_{1}$ as well as aperture type that give the minimum scatter. In these fits, we fix the planet's orbital parameters to the most recent values in the literature: $P = 2.2437661$~d, $i=84.32^{\circ}$, $a/R_{*}= 5.93$, $e=0.0822$, and $\omega = 251.67^{\circ}$ \citep{joshi,knutson2014}. For the 3.6~$\mu$m data set, we find that a fixed aperture with a radius of 1.8~pixels and $r_{0}=3.0$ produce the minimum scatter. When using a fixed aperture, the noise pixel parameter is not needed, so $r_{1}$ is undefined. In the 4.5~$\mu$m bandpass, we prefer a fixed aperture with a radius of 2.9~pixels and $r_{0}=3.5$.

Prior to fitting the selected photometric series with our full light curve model, we use a moving median filter to iteratively remove points with measured fluxes, $x$ positions, $y$ positions, or $\sqrt{\widetilde{\beta}}$ values that vary by more than 3$\sigma$ from the corresponding median values in the adjacent 64 frames in the time series. Choosing a larger or smaller interval for computing the median values does not significantly affect the number of excised points. The percentages of excised points are 1.8$\%$ and 1.6$\%$ in the 3.6 and 4.5~$\mu$m bandpasses, respectively.

\section{Data analysis}\label{sec:analysis}

\subsection{Transit and secondary eclipse model}\label{subsec:eventmodel}

Each full-orbit observation contains one transit and two secondary eclipses. We model these events using the formalism of \citet{mandelagol}. The transit light curve includes four free parameters: the scaled orbital semi-major axis $a/R_{*}$, the inclination $i$, the center of transit time $t_{T}$, and the planet-star radius ratio $R_{p}/R_{*}$, which is the square root of the relative transit depth. Each secondary eclipse event is defined by a center of eclipse time $t_{E}$ and a relative eclipse depth $d$, measured with respect to the value of the phase curve at mid-transit, in addition to $a/R_{*}$ and $i$, thus yielding four additional free parameters: $t_{E1}$, $t_{E2}$, $d_{1}$, and $d_{2}$. We ensure continuity between the phase curve and secondary eclipse light curves by scaling the amplitudes of eclipse ingress and egress (when the planet is partially occulted by the star) appropriately to match the out-of-eclipse phase curve values at the start and end of the eclipse. The host star WASP-14 has an effective temperature $T_{*} = 6462 \pm 75$~K, a specific gravity $\log{g} = 4.29\pm 0.04$, and a metallicity of [Fe/H] $=-0.13\pm0.08$ \citep{torres}. We model the limb-darkening in each bandpass using a four-parameter non-linear limb-darkening law with parameter values calculated as described in \citet{sing} for a 6500~K star with  $\log{g}=4.50$ and [Fe/H] $= -0.10$: $c_{1}-c_{4}=[-0.0192,0.7960,-0.8558,0.2983]$ at 3.6~$\mu$m and $c_{1}-c_{4}=[0.0225,0.3828,-0.2748,-0.0522]$ at 4.5~$\mu$m.\footnote[1]{Tables of limb-darkening parameter values, calculated in the \textit{Spitzer} bandpasses for various stellar temperatures, specific gravities, and metallicities, can be found on David Sing's website: www.astro.ex.ac.uk/people/sing}

\subsection{Phase curve model}\label{subsec:phasemodel}
 WASP-14b has a relatively low eccentricity of 0.08, and therefore the variation in the planet's apparent brightness throughout an orbit can be modeled to first order as a simple sinusoidal function of the true anomaly $f$ \citep{lewis}, analogous in form to a simple sine or cosine of the orbital phase angle that is used in the case of a circular orbit \citep{cowanagol}:
 \begin{equation}\label{phase}F(t)=F_{0}+c_{1}\cos(f(t)-c_{2}), \end{equation}
where $F_{0}$ is the star's flux (assumed to be constant and normalized to one), $c_{1}$ is the amplitude of the phase variations, and $c_{2}$ represents the lag between the peak of the planet's temperature and the time of maximum incident stellar flux due to the finite atmospheric radiative timescale of the planet. Here, $c_{1}$ and $c_{2}$ are free parameters that are computed in our fits.  We also experimented with other functional forms of the phase curve that included higher harmonics, but all of them resulted in higher values of the Bayesian Information Criterion (see Section~\ref{subsec:ramp} for more information).

\subsection{Correction for intrapixel sensitivity variations}\label{subsec:correction}
Photometric data obtained using \textit{Spitzer}/IRAC in the 3.6 and 4.5~$\mu$m bandpasses exhibit a well-studied instrumental effect due to intrapixel sensitivity variations \citep{charbonneau2005}. Small changes in the telescope pointing during observation cause variations in the measured flux from the target, resulting in a characteristic sawtooth pattern in the raw extracted photometric series. In our analysis, we decorrelate this instrumental systematic in two ways.

Our first approach to removing the intrapixel sensitivity effect is called pixel mapping \citep[][]{ballard,lewis}. In an image $j$, the location of the target on the array is given by the measured centroid position $(x_{j},y_{j})$, and the sensitivity of the pixel at that location is determined by comparing other images with measured centroid positions near $(x_{j},y_{j})$. The effective pixel sensitivity at a given position is calculated as follows:
\begin{align}\label{pixelmap}F_{\mathrm{meas},j}=F_{j}\sum\limits_{i=0}^{m}&e^{-(x_{i}-x_{j})^2/2\sigma_{x,j}^2}\times e^{-(y_{i}-y_{j})^2/2\sigma_{y,j}^2}\\ \notag
&\times e^{-\left(\sqrt{\widetilde{\beta}_{i}}-\sqrt{\widetilde{\beta}_{j}}\right)^2/2\sigma_{\sqrt{\widetilde{\beta}},j}^2}.\end{align}
Here, $F_{\mathrm{meas},j}$ is the flux measured in the $j$th image and $F_{j}$ is the intrinsic flux; $x_{j}$, $y_{j}$, and $\widetilde{\beta}$ are the measured $x$ position, $y$ position, and noise pixel parameter values. The quantities $\sigma_{x,j}$, $\sigma_{y,j}$, and $\sigma_{\sqrt{\widetilde{\beta}},j}$ are the standard deviations of $x$, $y$, and $\sqrt{\widetilde{\beta}}$ over the full range in $i$. For each image $j$, the summation in Eq.~\eqref{pixelmap} runs over the nearest $m=50$ neighbors of the stellar target, where we define distance as 
\begin{equation}\label{distance}d^{2}_{i,j} = (x_{i}-x_{j})^2+(y_{i}-y_{j})^2+\left(\sqrt{\widetilde{\beta}_{i}}-\sqrt{\widetilde{\beta}_{j}}\right)^2.\end{equation}
This method in effect adaptively smoothes the raw pixel map, allowing for a finer spatial scale in regions where the density of points is high while using a coarser spatial scale in regions with sparser sampling. We chose this number of neighbors to be large enough to adequately map the pixel response while maintaining a reasonably low computational overhead \citep{lewis}. Several previous studies of \textit{Spitzer} phase curves \citep[e.g.,][]{knutson2012, zellem} do not include the noise pixel parameter term in Eq.~\eqref{pixelmap}; we find that for our WASP-14b data, including the noise pixel parameter term produces $\sim5-10\%$ smaller residual scatter from our best-fit light curve solution in both bandpasses.

\begin{figure}[t*]
\begin{center}
\includegraphics[width=9cm]{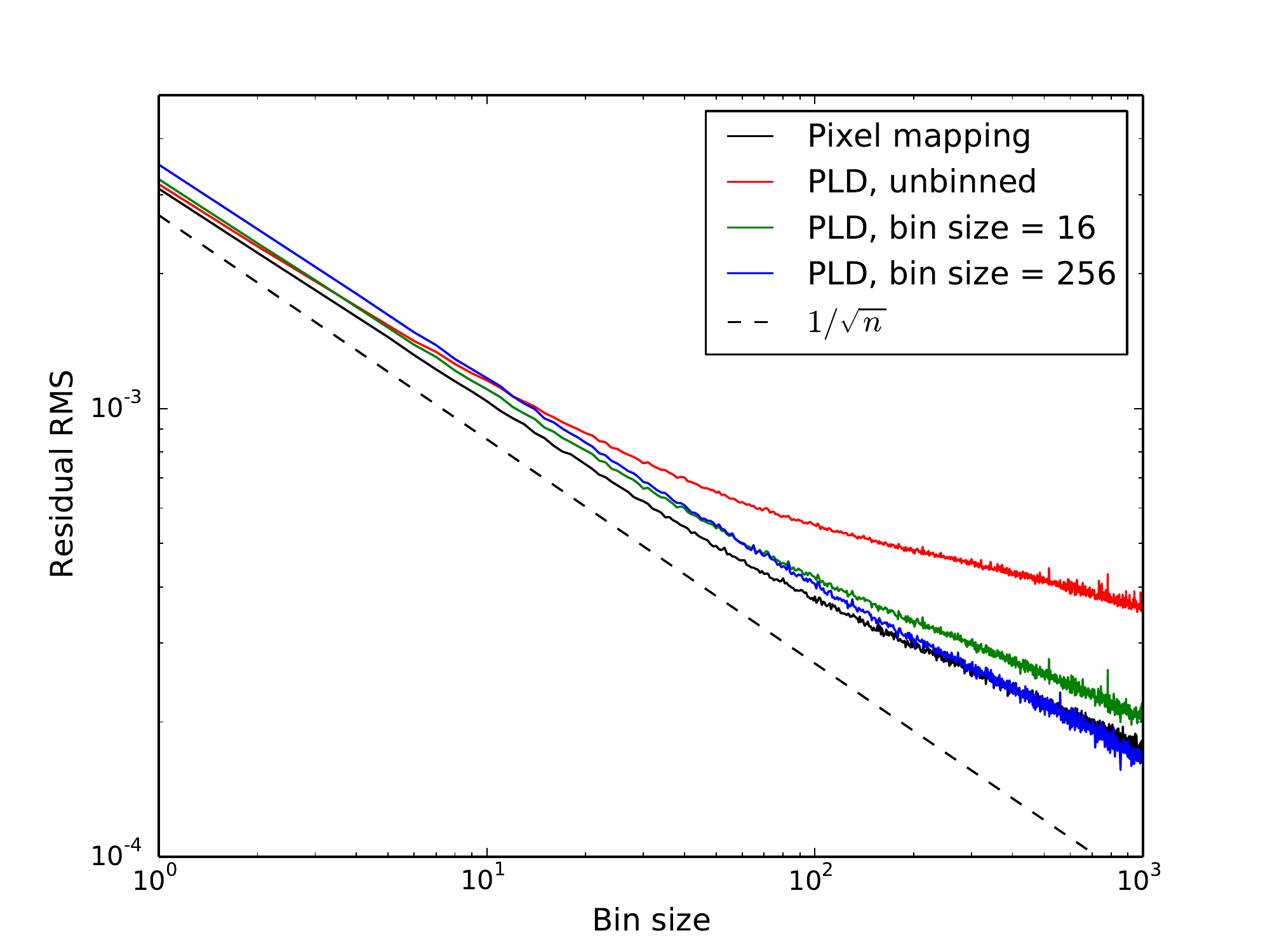}
\end{center}
\caption{Plot comparing the binned residual RMS resulting from the fit of the 3.6~$\mu$m photometric series to the light curve model using various types of instrumental noise decorrelation: pixel mapping (solid black line) and pixel level decorrelation with no binning (solid red line), 16-point bins (solid green line), and 256-point bins (solid blue line). For comparison, the $1/\sqrt{n}$ dependence of white noise on bin size is shown by the dashed black line;the trend is normalized at bin size $n=1$ to the photon noise level corresponding to the median photon count over the 3.6~$\mu$m photometric series, with sky background included.} \label{comparison}
\end{figure}

The second approach is a recently proposed technique known as pixel-level decorrelation \citep[PLD; see][for a complete description]{pld}. Unlike most other treatments of the intrapixel sensitivity effect, PLD does not attempt to relate the variations in the calculated position of the target on the pixel to the apparent intensity fluctuations. Rather, it utilizes the actual measured intensities of the individual pixels spanning the stellar PSF to provide an expression of the total measured flux. We consider pixels lying in a $3\times 3$ box centered on the star, which have pixel intensities $P_{k}(t)$, $k=1,\dots 9$. We divide each 3 $\times$ 3 pixel box by the summed flux over all nine pixels in order to remove (at least to zeroth order) any astrophysical flux variations, giving the following relation:
\begin{equation}\label{normalized}\hat{P}_{k}(t) = \frac{P_{k}(t)}{\sum_{k=1}^{9}P_{k}(t)}.\end{equation}
The intrapixel sensitivity effect is modeled as a linear combination of the arrays $\hat{P}_{k}$, and thus the total measured intensity $S$ is given by
\begin{equation}\label{pldtotal}S(t) = \left(\sum\limits_{k=1}^{9}b_{k}\hat{P}_{k}(t)\right) + b_{t}t + F(t) + h,\end{equation}
where $F(t)$ is the astrophysical model, comprising the phase curve, transit, and eclipses. The parameters $b_{k}$ are the linear coefficients that are determined through least-squares fitting, and $h$ is a free normalization parameter that corrects for the overall numerical offset introduced by the linear sum of pixel intensity arrays. Following \citet{pld}, we also include a linear ramp in time with a slope parameter $b_{t}$.

In \citet{pld}, PLD was applied to fitting \textit{Spitzer} secondary eclipses of several exoplanets, including WASP-14b. PLD was found to be generally more effective in removing time-correlated (i.e., red) noise, resulting in lower residual scatter from the best-fit solution when compared with other decorrelation techniques. The best results were obtained when the photometric time series were binned, pixel by pixel, prior to PLD fitting, since binning improves the precision of the measured intensities in pixels at the edge of the stellar PSF and can be adjusted to reduce the noise on the timescale of interest. When fitting our full-orbit WASP-14b photometric series using the PLD technique, we experimented with fitting either unbinned or binned data, with bin sizes equal to powers of two up to 256 ($\sim$8.5 minutes). We found that larger bin sizes, comparable to or exceeding the occultation ingress/egress timescale of roughly 20 minutes, cause excessive loss of temporal resolution and yielded unsatisfactory secondary eclipse and transit light curve fits. After fitting the model light curve to the data using PLD, we subtract the best-fit solution from the raw unbinned data to produce the residual series, which we use to evaluate the relative amount of time-correlated noise remaining in the data.

In Figure~\ref{comparison}, we compare the noise properties of each version of the 3.6~$\mu$m residual time series to the ideal $1/\sqrt{n}$ scaling we would expect for the case of independent (i.e., ``white" noise) Gaussian measurement errors, where $n$ is the bin size. The white noise trend is normalized at bin size $n=1$ to the photon noise level corresponding to the median photon count over the observation data set (with sky background included). Comparing the various PLD fits, we find that at small bin sizes, unbinned PLD gives the lowest residual scatter, while at larger bin sizes approaching the duration of eclipse ingress or egress, PLD with larger bins results in lower residual scatter. The same trend is seen when fitting the 4.5~$\mu$m data. \citet{pld} found that the optimal PLD performance is achieved when the range of star positions is lower than 0.2~pixels. The range of pixel motion in our full-orbit phase curve observations modestly exceeds this limit, and we indeed find that the residual scatter from the pixel mapping fits is smaller than the scatter from any of the PLD fits. We conclude that the pixel mapping technique produces the lowest residual scatter for our full-orbit observation data sets, and we therefore use this technique in the final version of our analysis.

In addition to having higher residual scatter than the pixel mapping solutions, the PLD fits yield eclipse depth and phase curve parameter estimates that often differ strongly from the corresponding values derived from fits with pixel mapping; these discrepancies sometimes exceed the $3\sigma$ level. The best-fit values derived from the PLD fits also display a higher level of variation across different choices of binning, photometric aperture, and exponential ramp type (see Section~\ref{subsec:ramp}) than in the case of pixel mapping fits. This points toward an inherent instability in the PLD method when fitting full-orbit phase curves that may be related to the larger range in star motions characteristic of such data sets.

\subsection{Exponential ramp correction}\label{subsec:ramp}
Previous studies using \textit{Spitzer}/IRAC have noted a short-duration ramp at the beginning of each observation, and again after downlinks \citep[e.g.][]{knutson2012,lewis}. The ramp has a characteristic asymptotic shape that typically decays to a constant value on timescales of an hour or less in the 3.6 and 4.5~$\mu$m bandpasses. We first experimented with removing the first 30 or 60 minutes of data from each phase curve observation, selecting the removal interval that minimizes the residual RMS from the best-fit solution binned in five-minute intervals. We find that in both bandpasses, we obtain the best results when we do not trim any data from the start of the observations. When examining the residual time series, we noticed a small ramp visible at the start of the 3.6~$\mu$m observations.  We therefore considered whether or not our fits might be further improved by the addition of an exponential function.

We experimented with including an exponential ramp in our phase curve model, using the formulation given in \citet{agol}:
\begin{equation}\label{ramp}F=1\pm a_{1}\mathrm{e}^{-t/a_{2}}\pm a_{3}\mathrm{e}^{-t/a_{4}},\end{equation}
where $t$ is the time since the beginning of the observation, and $a_{1}$---$a_{4}$ are correction coefficients. To determine whether this function is necessary and if so, how many exponential terms to include in the ramp model, we use the Bayesian Information Criterion (BIC), defined as
\begin{equation}\label{bic}\mathrm{BIC} = \chi^2 + k\ln{N},\end{equation}
where $k$ is the number of free parameters in the fit, and $N$ is the number of data points. By minimizing the BIC, we select the type of ramp model that yields the smallest residuals without  ``over-fitting'' the data. For the 3.6~$\mu$m data, we find that using a single exponential ramp gives a marginally lower BIC compared to the no-ramp case, while for the 4.5~$\mu$m data, no ramp is needed at all. The residuals from the best-fit full phase curve solution, shown in Figures~\ref{phase1} and \ref{phase2}, do not appear to display any uncorrected ramp-like behaviors.

\subsection{Parameter fits}\label{subsec:fits}

\begin{table*}[t!]
\centering
\begin{threeparttable}
\caption{Best-fit Parameters} \label{tab:values}

\renewcommand{\arraystretch}{1.2}
\begin{center}
\begin{tabular}{m{0.001cm} l m{0.001cm} r m{0.001cm} r m{0.001cm} }
\hline\hline
& \multicolumn{1}{c}{Parameter}& & \multicolumn{1}{c}{3.6~$\mu$m} & &  \multicolumn{1}{c}{4.5~$\mu$m} & \\
\hline\\
& \textit{Transit Parameters} & & & & &\\
& $R_{p}/R_{*}$ & &$0.09416^{+0.00057}_{-0.00068}$ & & $0.09421^{+0.00047}_{-0.00070}$ & \\
& $t_{T}$~(BJD)\textsuperscript{a} & & $2456034.21228^{+0.00023}_{-0.00026}$ & &  $2456043.18707^{+0.00026}_{-0.00025}$ & \\

\\
& \textit{Eclipse Parameters} & & & & &\\
& 1\ts{st} eclipse depth, $d_{1}$~($\%$) & & $0.1859^{+0.0096}_{-0.0108}$ & & $0.2115^{+0.0135}_{-0.0114}$ & \\
& $t_{E1}$~(BJD)\textsuperscript{a}\textsuperscript{,}\textsuperscript{b} & & $2456033.05277^{+0.00092}_{-0.00101}$ & & $2456042.02887^{+0.00091}_{-0.00080}$ & \\
& 2\ts{nd} eclipse depth, $d_{2}$~($\%$)\textsuperscript{$\dag$} & & $0.1889^{+0.0060}_{-0.0049}$ & & $0.2367^{+0.0096}_{-0.0142}$ & \\
& $t_{E2}$~(BJD)\textsuperscript{a}\textsuperscript{,}\textsuperscript{b}\textsuperscript{,}\textsuperscript{$\dag$} & & $2456035.29948^{+0.00057}_{-0.00045}$ & & $2456044.27400^{+0.00084}_{-0.00072}$ & \\

\\
& \textit{Orbital Parameters} & & & & &\\
& Inclination, $i$~($^{\circ}$) & & $84.65^{+0.35}_{-0.36}$ & & $84.61^{+0.33}_{-0.34}$ &\\
& Scaled semi-major axis, $a/R_{*}$ & & $6.01^{+0.14}_{-0.13}$  & & $5.98\pm 0.13$ &\\

\\
& \textit{Phase Curve Parameters} & & & & &\\
& Amplitude, $c_{1}$~($\times 10^{-4}$) & & $9.70^{+0.38}_{-0.39}$ & & $7.86^{+0.22}_{-0.24}$ &\\
& Phase shift, $c_{2}$~($^{\circ}$) & & $5.7^{+2.4}_{-2.2}$ & & $9.4^{+2.5}_{-2.3}$ &\\
& Maximum flux offset~(h)\textsuperscript{c} & & $-1.43\pm 0.21$  & & $-1.01\pm 0.21$ &\\
& Minimum flux offset~(h)\textsuperscript{c} & &$-2.03^{+0.42}_{-0.39}$  & & $-1.39^{+0.43}_{-0.40}$ &\\

\\
& \textit{Ramp Parameters} & & & & &\\
& $a_{1}$ ($\times 10^{-4}$) & & $-4.7^{+1.8}_{-2.0}$ & & --- &\\
& $a_{2}$ (d)& & $0.106^{+0.113}_{-0.054}$ & & --- &\\

\\
\hline\hline
\end{tabular}

\begin{tablenotes}
      \small
      \item \textsuperscript{a}All times are listed in BJD$_{\mathrm {UTC}}$ for consistency with other studies; to convert to BJD$_{\mathrm {TT}}$, add 66.184~s \citep{eastman}.
      \item \textsuperscript{b}The center of secondary eclipse times are not corrected for the light travel time across the system, $\Delta t= 35.9$~s.
      \item \textsuperscript{c}The maximum and minimum flux offsets are measured relative to the center of secondary eclipse time and center of transit time, respectively, and are derived from the phase curve fit parameters $c_{1}$ and $c_{2}$. The maximum flux offset reported is the error-weighted mean of the flux offsets relative to the first and second secondary eclipses.
       \item \textsuperscript{$\dag$}These values are computed from fitting the second 3.6~$\mu$m secondary eclipse separately using the pixel-level decorrelation (PLD) method. This was done in order to remove an anomalous signal that occurs during the eclipse in the global 3.6~$\mu$m phase curve fit using pixel mapping.

    \end{tablenotes}
    \end{center}
    \end{threeparttable}
\end{table*}

\begin{figure}[t!]
\begin{center}
\includegraphics[width=9cm]{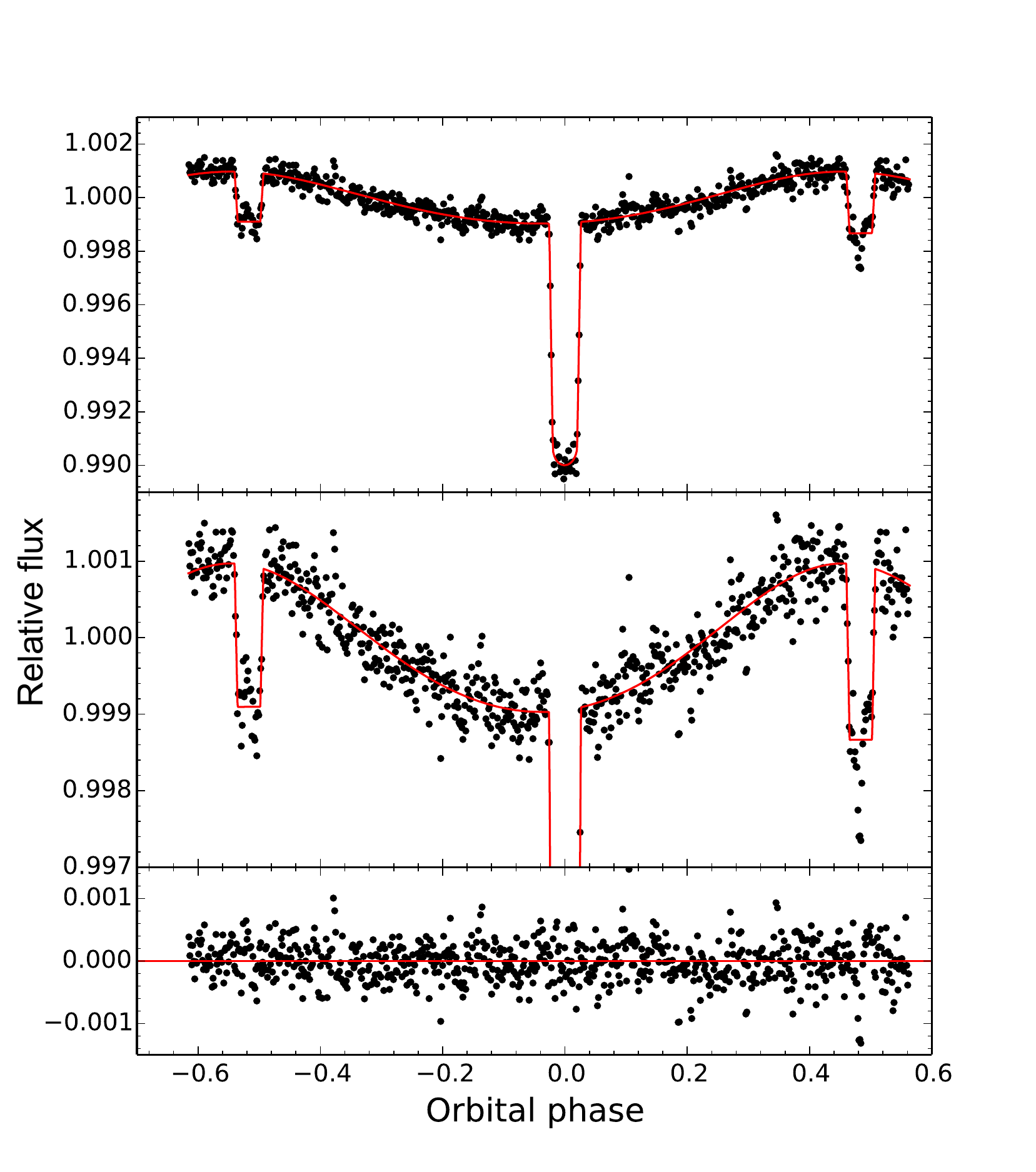}
\end{center}
\caption{Top panel: Final 3.6~$\mu$m photometric series with instrumental variations removed, binned in five-minute intervals (black dots). The best-fit total phase, transit, and eclipse light curve is overplotted in red. Middle panel: The same data as the upper panel, but with an expanded y axis for a clearer view of the phase curve. Bottom panel: The residuals from the best-fit solution.}\label{phase1}
\end{figure}

\begin{figure}[t!]
\begin{center}
\includegraphics[width=9cm]{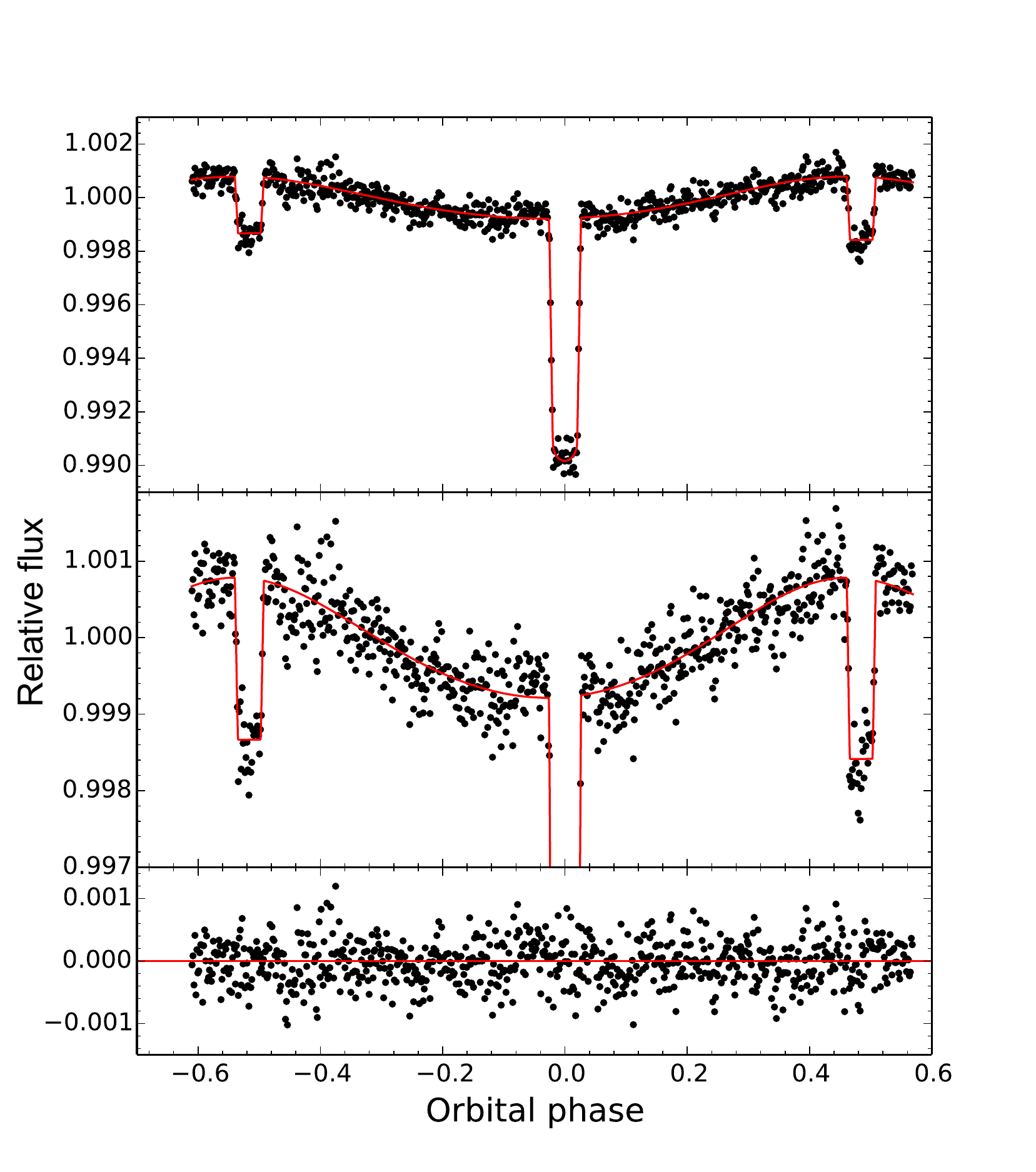}
\end{center}
\caption{Final 4.5~$\mu$m photometric series with instrumental variations removed and the corresponding residual series; see Figure~\ref{phase1} for a complete description.} \label{phase2}
\end{figure}

\begin{figure}[t*]
\begin{center}
\includegraphics[width=9cm]{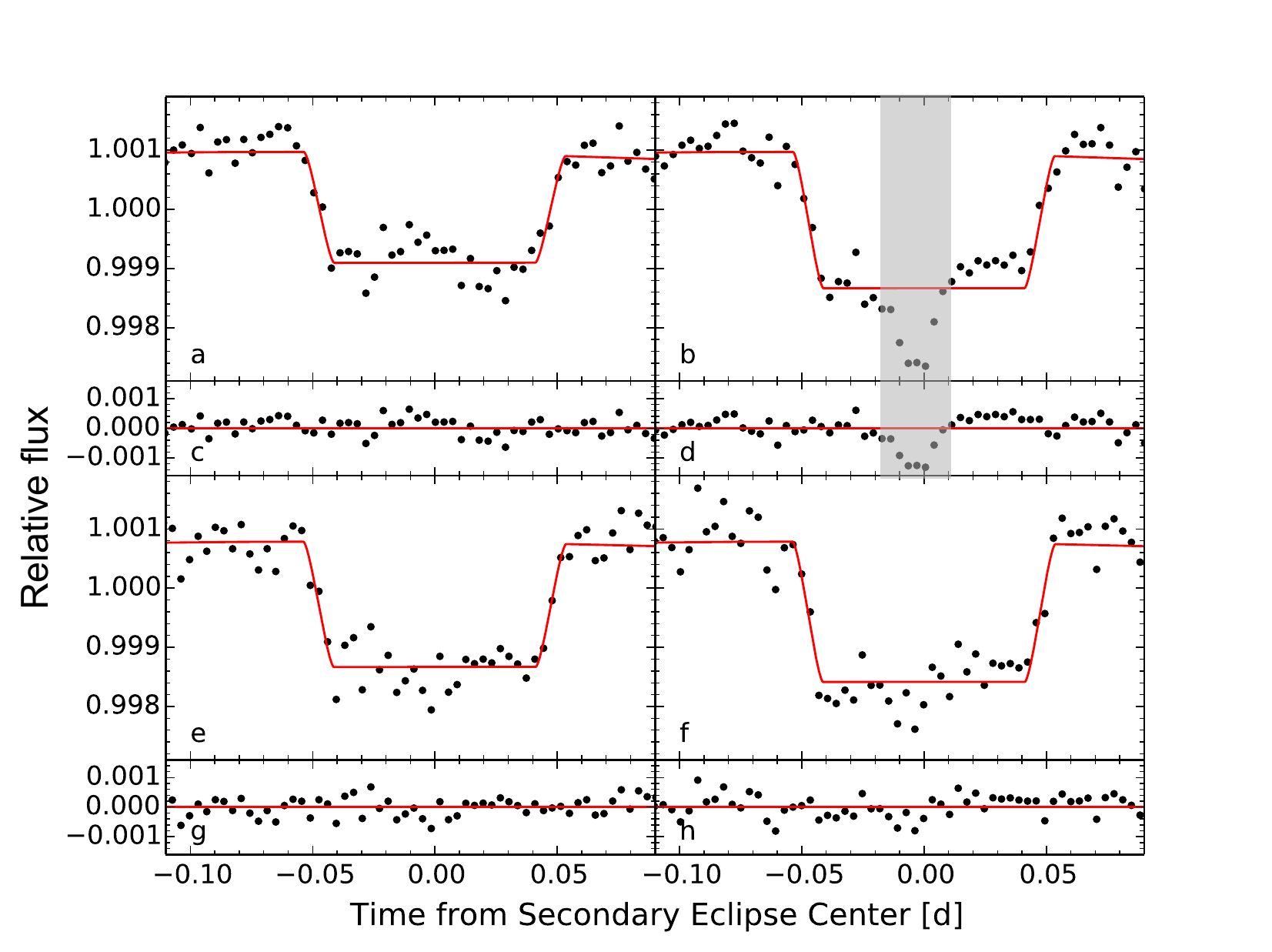}
\end{center}
\caption{Best-fit eclipse light curve data in the 3.6~$\mu$m (a$-$d) and 4.5~$\mu$m (e$-$h) bands after correcting for intrapixel sensitivity variations, binned in five-minute intervals (black dots). The best-fit model light curves are overplotted in red. The residuals from the best-fit solution (c$-$d, g$-$h) are shown directly below the corresponding light curve data (a$-$b; e$-$f). In panels b and d, the anomalous residual signal (see text) is highlighted by the gray box.} \label{eclipses}
\end{figure}

\begin{figure}[t*]
\begin{center}
\includegraphics[width=9cm]{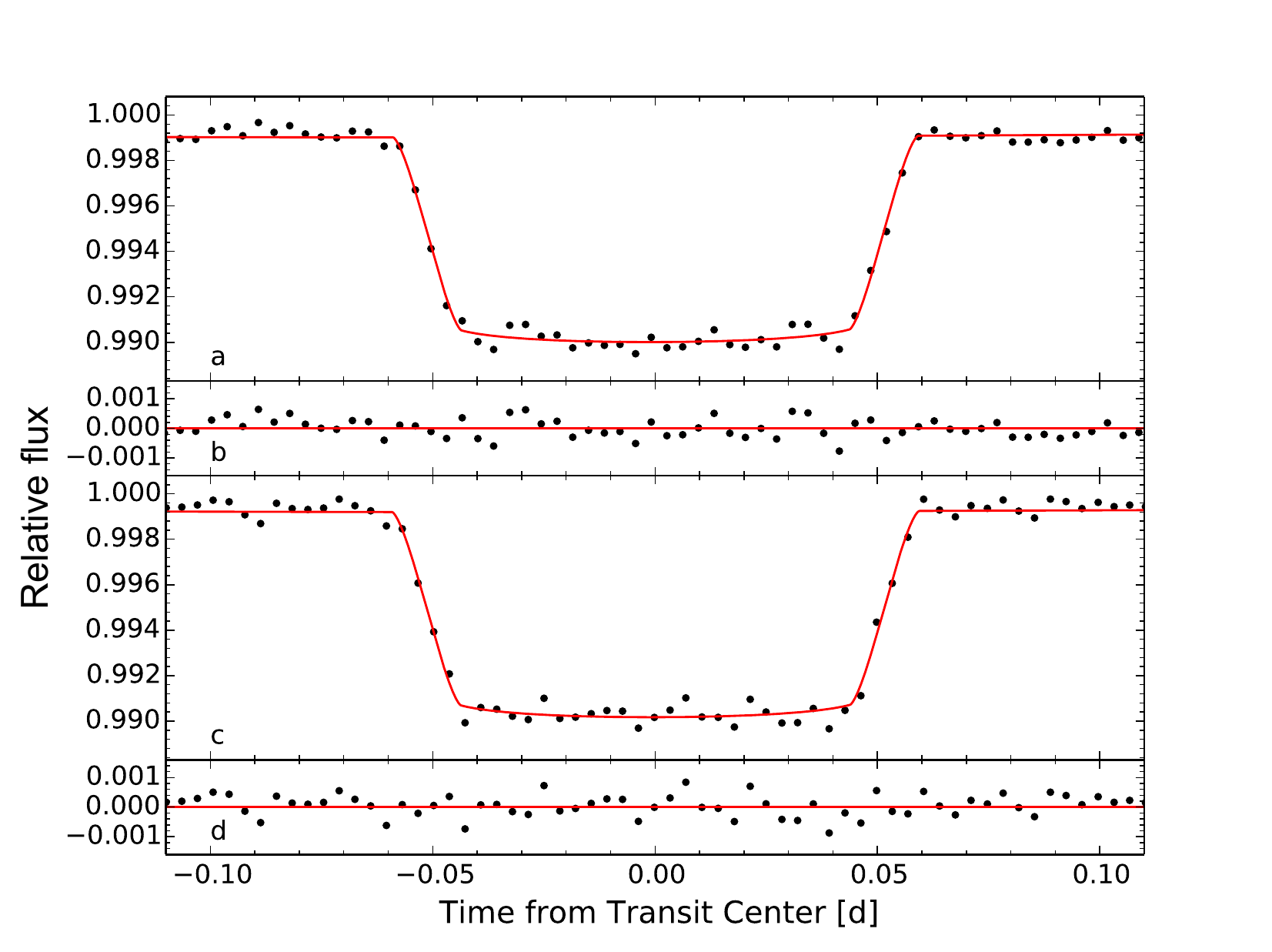}
\end{center}
\caption{Best-fit transit light curve data in the 3.6~$\mu$m and 4.5~$\mu$m bands after correcting for intrapixel sensitivity variations, binned in five-minute intervals (black dots). The best-fit model light curves are overplotted in red. The residuals from the best-fit solution are shown directly below the corresponding light curve data.} \label{transits}
\end{figure}

We use a Levenberg-Marquardt least-squares algorithm to fit each full-orbit photometric series to our total model light curve, with the intrapixel sensitivity correction calculated via pixel mapping. In the final version of these global fits, we use the updated values for $e$ and $\omega$ obtained from our radial velocity analysis (see Section~\ref{subsec:ephem}) as well as the updated orbital period calculated from fitting all published transits. The best-fit transit, secondary eclipse, and phase parameters are listed in Table~\ref{tab:values} along with their uncertainties. Figures~\ref{phase1} and \ref{phase2} show the full-orbit data in the 3.6 and 4.5~$\mu$m bandpasses, respectively, with instrumental variations removed. The individual eclipse and transit light curves are shown in Figures~\ref{eclipses} and \ref{transits}.

In the 3.6~$\mu$m light curve data, there is an anomalous signal in the residuals from the global phase curve fit that occurs during the last secondary eclipse. This anomaly is characterized by a short $\sim$20-minute dip in the middle of the eclipse and was not removed by the pixel mapping method, resulting in a $3.1\sigma$ discrepancy between the eclipse depths from the first and second events. Similar short-duration anomalies (both positive and negative) have been reported in \textit{Spitzer} 3.6~$\mu$m data and are usually attributed to variations in the width of the stellar target \citep[e.g.,][]{lanotte}. Our pixel mapping technique accounts for these variations by incorporating the noise-pixel parameter pixel sensitivity calculation in Eq.~\eqref{pixelmap}, so we conclude that the anomaly in our data is likely attributable to some other instrumental effect. 

\begin{figure}[t*]
\begin{center}
\includegraphics[width=9cm]{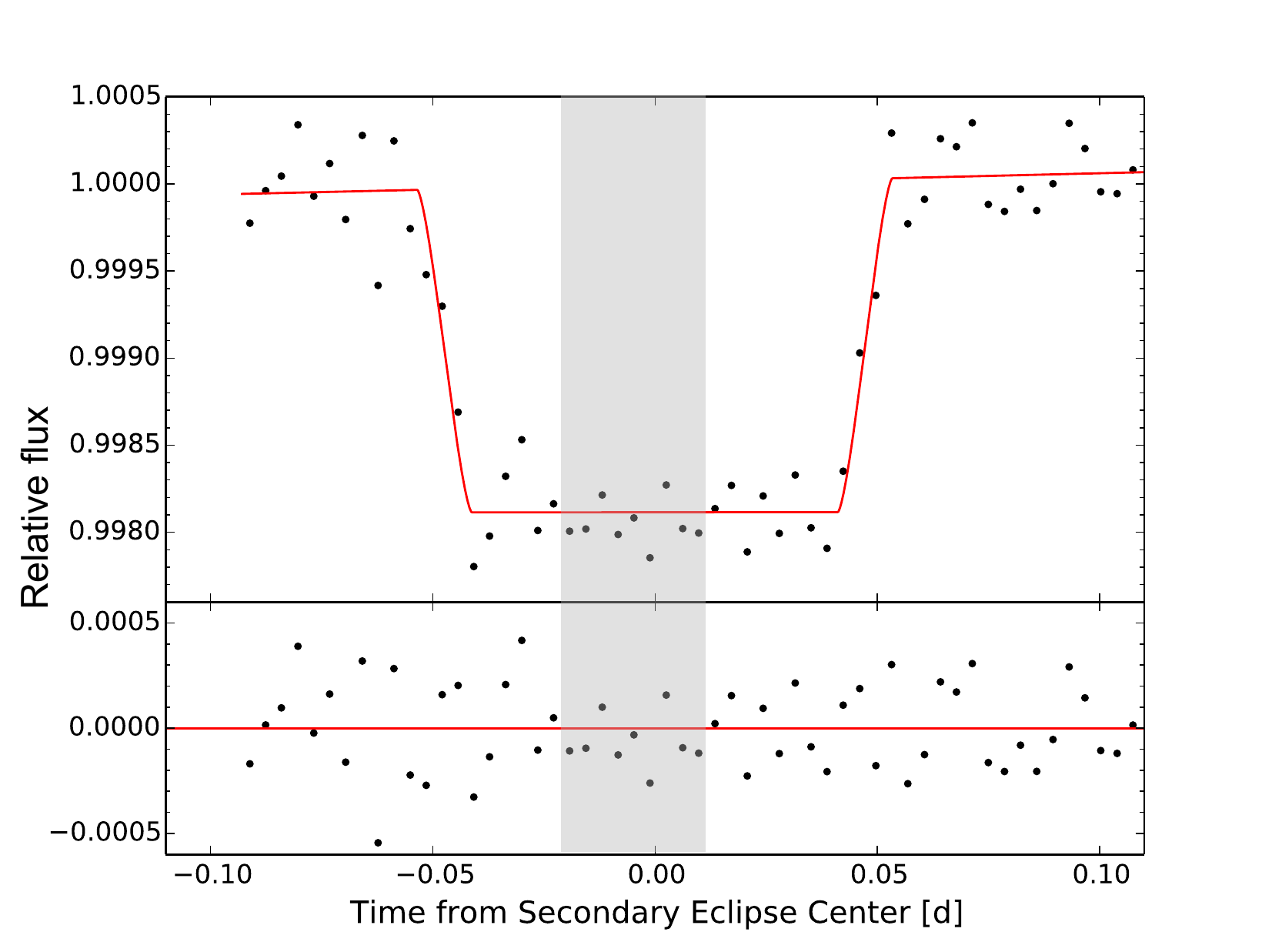}
\end{center}
\caption{Best-fit secondary eclipse light curve data for the second 3.6~$\mu$m eclipse, derived from fitting the eclipse event separately using PLD  (black dots). The data has been corrected for intrapixel sensitivity variations and is binned in five-minute intervals. The best-fit secondary eclipse model light curve is overplotted in red. The residuals from the best-fit solution are shown in the bottom panel. The gray box indicates the previous location of the anomalous signal (see text), which has been removed by the PLD method.} \label{eclipsefix}
\end{figure}

In order to recover the second 3.6~$\mu$m eclipse depth, we experimented with fitting the eclipse event separately. Selecting a short $\sim$0.2-day segment of the phase curve observation surrounding the eclipse, we fit the data to a simplified secondary eclipse light curve model using both pixel mapping and PLD. This model has three free parameters --- the center of eclipse time $t_{E2}$, the eclipse depth $d_{2}$, and a linear slope $c_{t}$, which accounts for the out-of-eclipse variation in the planet's brightness. The inclination, scaled semi-major axis, and planet-star radius ratio are fixed at the values derived from the global 3.6~$\mu$m phase curve fit and listed in Table~\ref{tab:values}. In these fits, we use the same choice of aperture as in the full phase curve fit; in the case of PLD, we optimize for the bin size based on the residual scatter and find that a bin size of 128 yields the lowest residual RMS. Comparing the results of our fits using pixel mapping and PLD, we see that while the residuals from the best-fit solution with pixel mapping show a significant anomalous signal similar to the one present in the global 3.6~$\mu$m phase curve fit, no residual anomaly is evident in the best-fit solution with PLD. The range of star positions during this segment of data is much less than 0.2~pixels, so we expect PLD to perform optimally. The best-fit parameter values from the PLD fit are $t_{E2}=2456035.29938^{+0.00056}_{-0.00045}$ (BJD$_{\mathrm {UTC}}$), $d_{2} = 0.1894^{+0.0059}_{-0.0049}~\%$, and $c_{t}=7.7^{+6.8}_{-6.9}\times 10^{-4}~\mathrm{d}^{-1}$. The data with instrumental effects removed and the best-fit secondary eclipse solution are shown in Figure~\ref{eclipsefix}. We note that the best-fit eclipse depth from our individual fit of the second 3.6~$\mu$m eclipse is consistent with the depth of the first 3.6~$\mu$m eclipse derived from the global phase curve fit at the $0.15\sigma$ level. Henceforth, we use the values from our individual secondary eclipse fit with PLD for the center of eclipse and eclipse depth of the second 3.6~$\mu$m eclipse.

Taking the error-weighted average of the eclipse depths listed in Table~\ref{tab:values} at each wavelength, we arrive at $0.1882\%\pm0.0048\%$ and $0.2247\%\pm 0.0086\%$ for the 3.6 and 4.5~$\mu$m bandpasses, respectively. These values are consistent with the ones reported in \citet{blecic} in their analysis of previous \textit{Spitzer} secondary eclipse observations to better than $1\sigma$ ($0.19\%\pm0.01\%$ at 3.6~$\mu$m and $0.224\%\pm0.018\%$ at 4.5~$\mu$m).

We estimate the uncertainties in our best-fit parameters in two ways. The first approach is the ``prayer-bead'' (PB) method \citep{gillon}, which gives an estimate of the contribution of time-correlated noise to the uncertainty. This method entails extracting the residuals from the best-fit solution, dividing the residuals into segments, and cyclically permuting the residual series segment by segment, each time adding the new residual series back to the best-fit solution and recomputing the parameters using the least-squares algorithm. For each free parameter, we create a histogram of the best-fit values from every permutation and calculate the uncertainties based on the $1\sigma$ upper and lower bounds from the median. 

The second approach is a Markov chain Monte Carlo (MCMC) routine with $10^5$ steps, where we initiate each chain at the best-fit solution from the least-squares analysis. The uncertainty on individual data points is set to be the standard deviation of the residuals from the best-fit solution. We discard an initial burn-in on each chain of length equal to 20$\%$ of the chain length, which we found ensured the removal of any initial transient behavior in a chain, regardless of the choice of initial state. The distribution of values for each parameter is close to Gaussian, and there are no significant correlations between pairs of parameters. As in the PB method, we set the uncertainties in the fitted parameters to be the $1\sigma$ upper and lower bounds from the median. For each parameter, we choose the larger of the two errors and report it in Table~\ref{tab:values}. We find that the PB errors are consistently larger and range between 1.0 and 2.7 times that of the corresponding MCMC errors.

\begin{figure}[t*]
\begin{center}
\includegraphics[width=8cm]{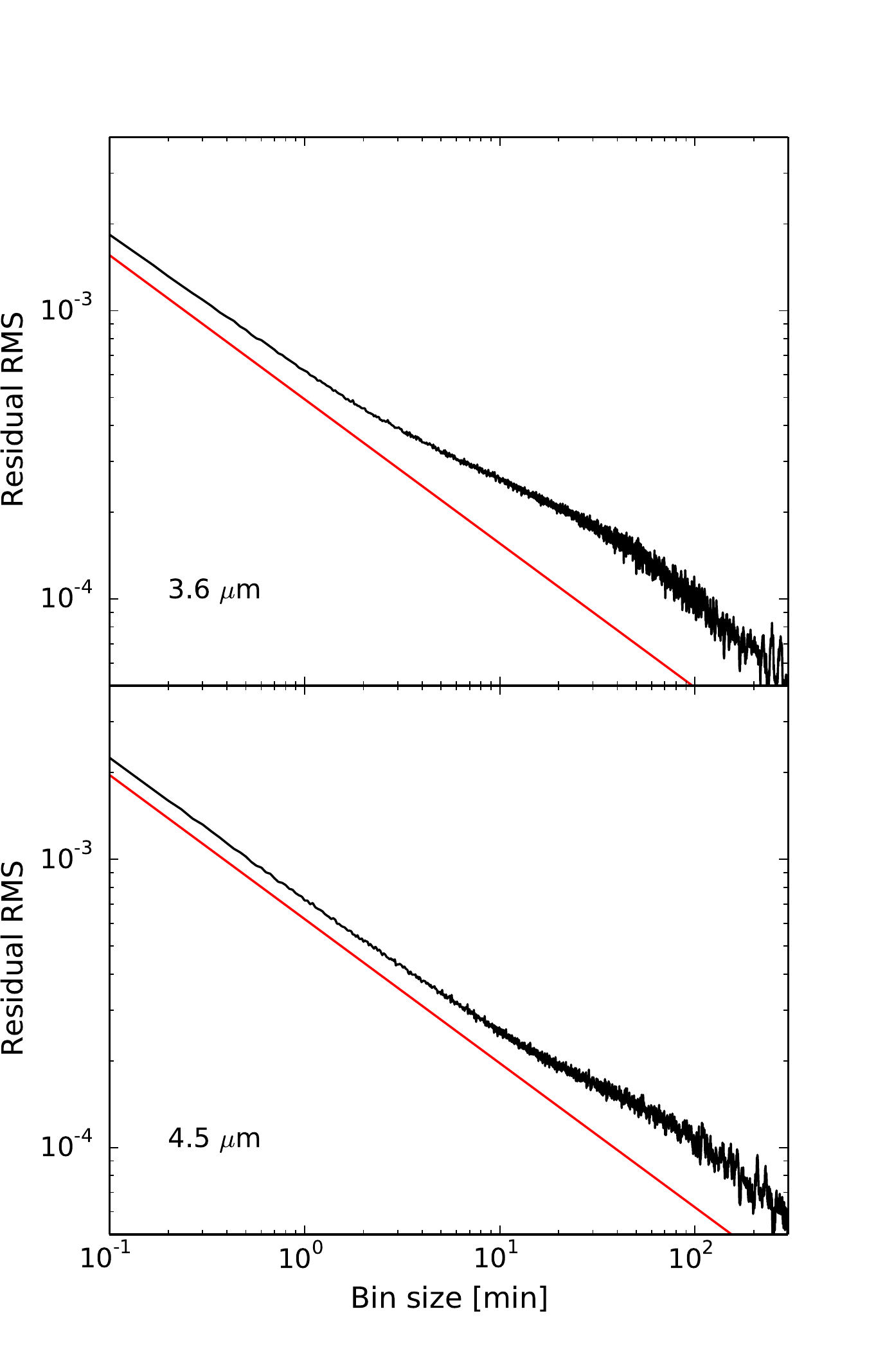}
\end{center}
\caption{Plot of the standard deviation of the residuals versus bin size for the 3.6~$\mu$m (top panel) and 4.5~$\mu$m (bottom panel) data sets after removing intrapixel sensitivity effects and dividing out the best-fit total light curve solutions (black lines). For comparison, the $1/\sqrt{n}$ dependence of white noise on bin size is shown by the red lines; the white noise trends are normalized to match the expected photon noise limit corresponding to the median photon count over each observation.}\label{rednoise}
\end{figure}

The RMS scatter in the best-fit residual series exceeds the predicted photon noise limit by a factor of 1.16 at 3.6~$\mu$m and 1.14 at 4.5~$\mu$m. We estimate the level of red noise by calculating the standard deviation of the best-fit residuals for various bin sizes, shown in Figure~\ref{rednoise} along with the inverse square-root dependence of white noise on bin size for comparison. On timescales relevant for the eclipses and transits (e.g., the ingress/egress timescale --- $\sim$15 minutes), the red noise increases the RMS by a factor of approximately 1.8 at 3.6~$\mu$m and approximately 1.4 at 4.5~$\mu$m.

\section{Discussion}\label{sec:dis}
\subsection{Orbital parameters and ephemeris}\label{subsec:ephem}
\begin{figure}[t*]
\begin{center}
\includegraphics[width=9cm]{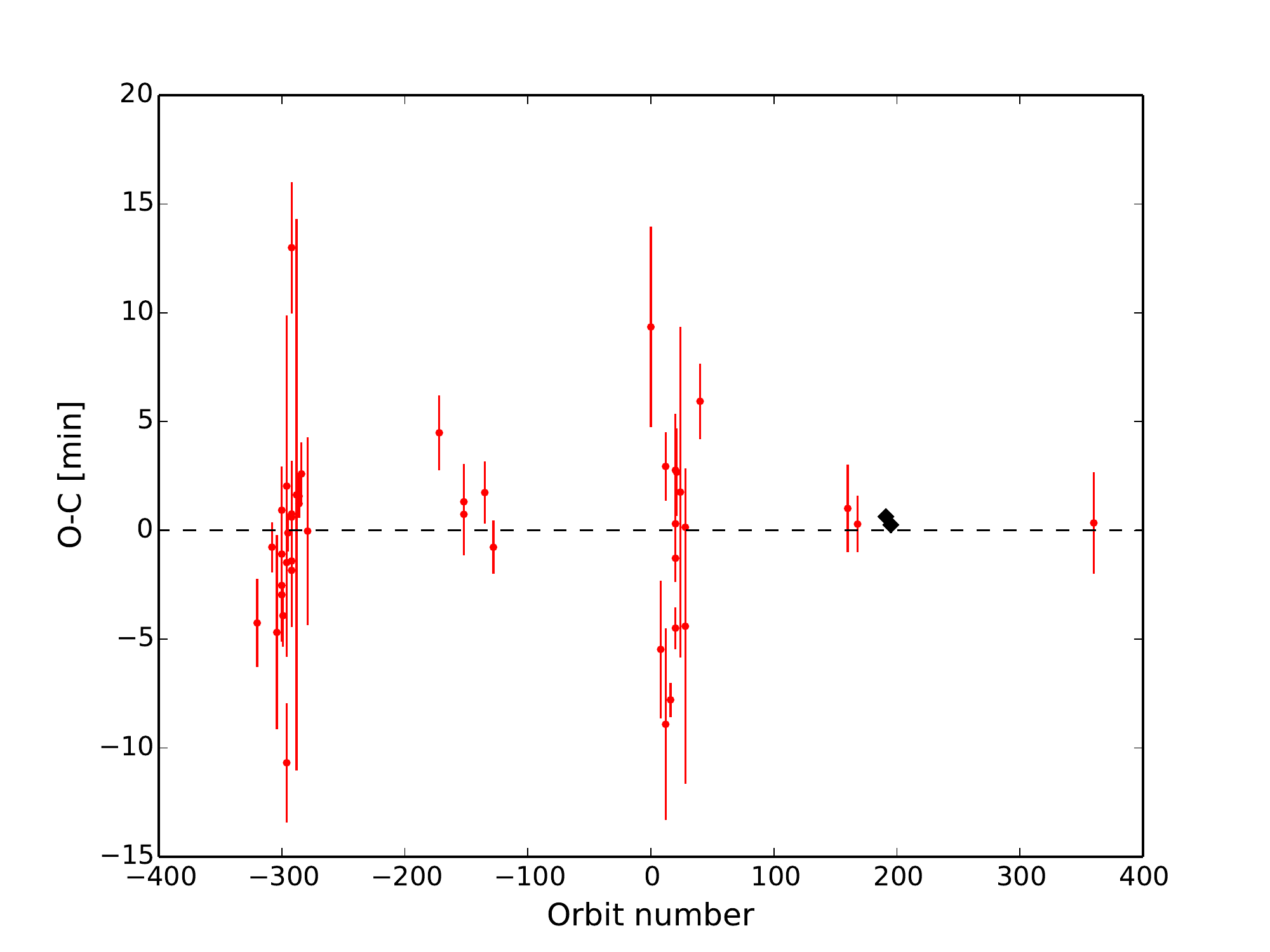}
\end{center}
\caption{Observed minus calculated transit times for all published observations \citep[red circles are previously-published values;][]{joshi,blecic,raetz} using the updated ephemeris calculated in Section~\ref{subsec:ephem}. The black diamonds are the two transit times measured from our phase curve data.}\label{transitoc}
\end{figure}

We combine the two transit times calculated from our phase curve observations with all other published values \citep{joshi,blecic,raetz} to arrive at an updated ephemeris for the WASP-14 system. Here, we define the zeroth epoch as that of the transit nearest in time to the error-weighted mean of all measured transit times. The transit observations span more than 5.3 years, and by fitting a line through the transit times, we derive new, more precise estimates of the orbital period $P$ and mid-transit time $T_{c,0}$:
\begin{equation}\label{ephem}\begin{cases} P &= 2.24376507 \pm 0.00000046~\mathrm{days}\\
T_{c,0} &= 2455605.65348 \pm 0.00011~(\mathrm{BJD}_{\mathrm{TDB}})\end{cases}.\end{equation}
Figure~\ref{transitoc} shows the observed minus calculated transit times derived from these updated ephemeris values.

\begin{figure}[t*]
\begin{center}
\includegraphics[width=9cm]{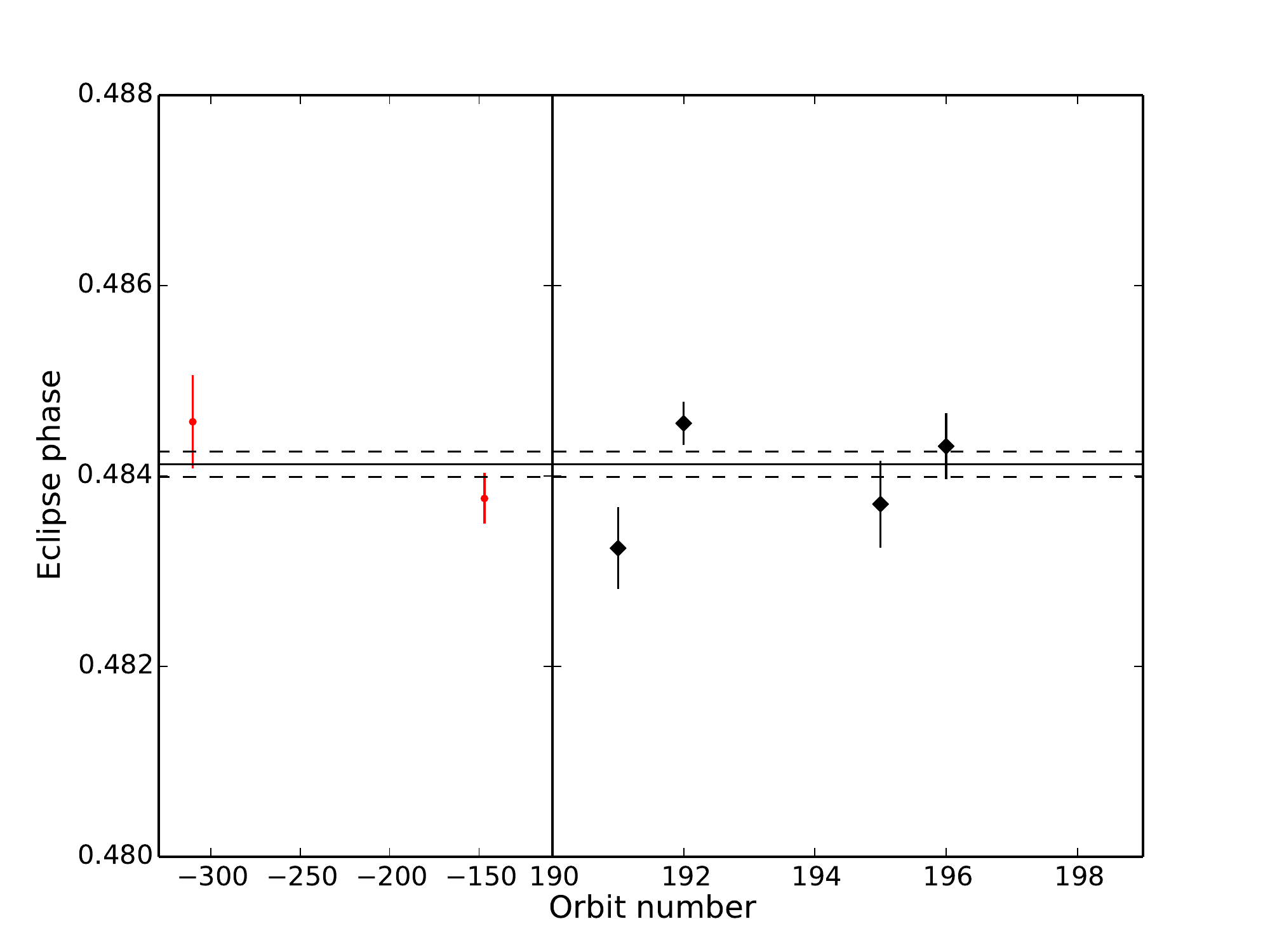}
\end{center}
\caption{Orbital phase of secondary eclipse for all published observations \citep[red circles are previously-published values from][]{blecic} using the updated ephemeris calculated in Section~\ref{subsec:ephem}. The black diamonds are the four secondary eclipse times measured from our phase curve data. The solid and dashed lines indicate the error-weighted mean phase value and corresponding $1\sigma$ confidence bounds, respectively. The horizontal axis has been condensed for clarity.}\label{eclipseoc}
\end{figure}

We use the secondary eclipse times to obtain a second, independent estimate of the orbital period. Carrying out a linear fit through the four secondary eclipse times calculated from our phase curve data and the two secondary eclipse times published in \citet{blecic}, we arrive at a best-fit value of $P=2.2437660\pm 0.0000017$~days. This period is consistent with the best-fit transit period at the $0.5\sigma$ level. In Figure~\ref{eclipseoc}, we plot the orbital phase of secondary eclipse for all published secondary eclipse times using the updated transit ephemeris values in Eq.~\eqref{ephem}. The error-weighted mean orbital phase of secondary eclipse is $0.48412\pm 0.00013$.

\LongTables
 \begin{deluxetable}{lrr}[t!]
     \tablecaption{Results from radial velocity fit \\ with priors on transit ephemeris and eclipse times}
    \tablehead{\colhead{Parameter} & \colhead{Value} & \colhead{Units}}
   
    \startdata
  
	\sidehead{RV Model Parameters}$P_{b}$ &  2.24376524 $\pm4.4\mathrm{E}-07$ & days\\
$T_{c,b}$ & 2456034.21261 $\pm 0.00015$ & \bjdtdb\\
$e_{b}$ & 0.0830 $^{+0.0029}_{-0.0030}$ & \\
$\omega_{b}$ & 252.67 $^{+0.70}_{-0.77}$ & degrees\\
$K_{b}$ & 986.4 $^{+2.6}_{-2.5}$ & \ms\\
$\gamma_{1}$ & 155.8 $\pm 2.6$ & \ms\\
$\gamma_{2}$ & $-$70.0 $^{+6.9}_{-6.6}$ & \ms\\
$\gamma_{3}$ & $-$73.7 $\pm 6.8$ & \ms\\
$\gamma_{4}$ & 187.4 $^{+7.3}_{-7.0}$ & \ms\\
$\dot{\gamma}$ & 0.0025 $\pm 0.0032$ & \ms day$^{-1}$\\
\sidehead{RV-derived Parameters}
$e\cos{\omega}$ & $-$0.02474 $^{+0.00078}_{-0.00074}$ & \\
$e\sin{\omega}$ & $-$0.0792 $^{+0.0031}_{-0.0029}$ & \\

 \enddata 
\tablenotetext{}{NOTE --- Radial velocity zero point offsets ($\gamma_{1-4}$) derived from four separate RV data sets: 1 -- Keck/HIRES \citep{knutson2014}, 2 -- FIES \citep{joshi}, 3 -- SOPHIE \citep{joshi}, 4 -- SOPHIE \citep{husnoo}. See text for description of other variables.}\label{tab:rv}
\end{deluxetable}

\begin{figure}[t!]
\begin{center}
\includegraphics[width=9cm]{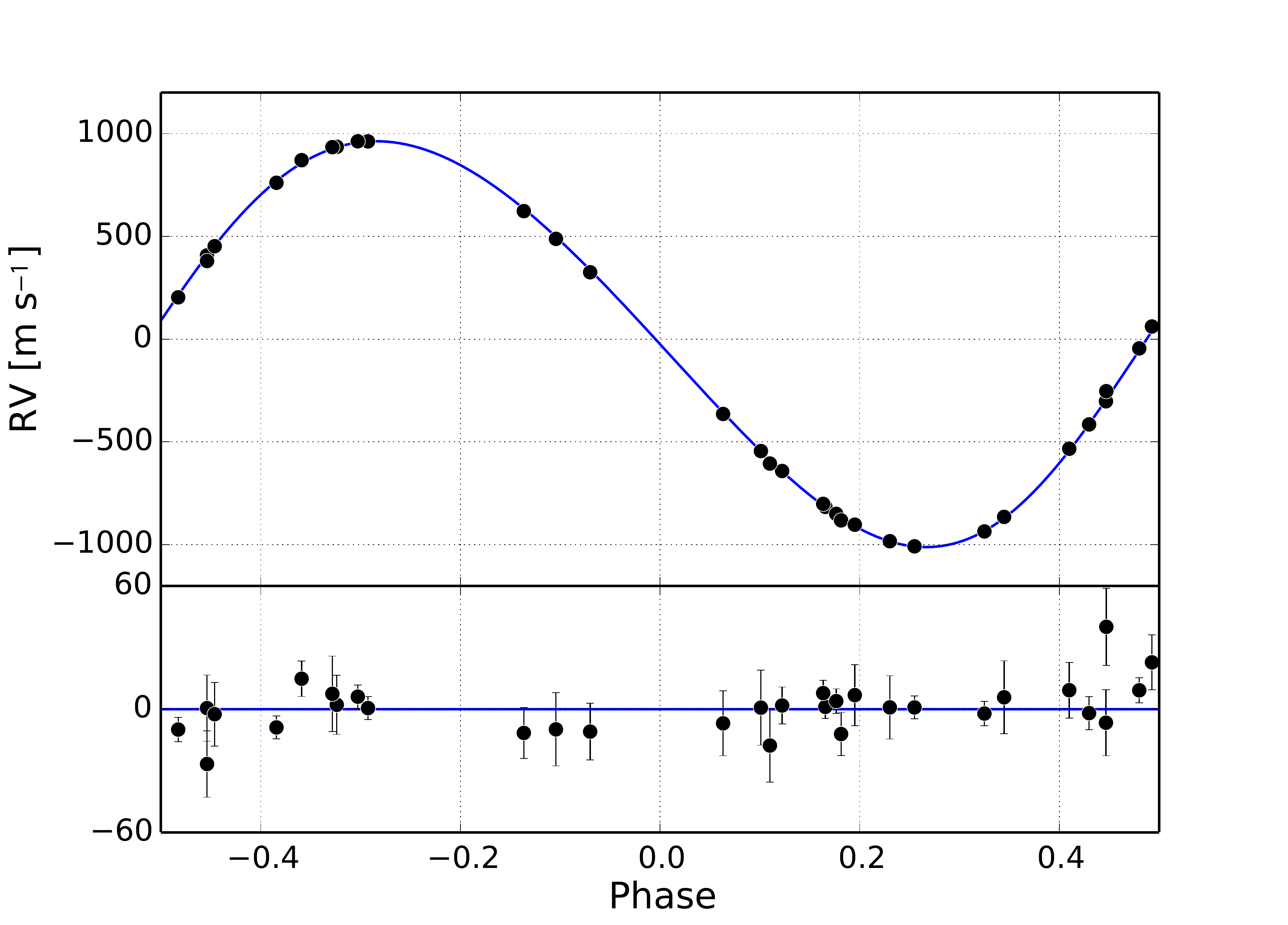}
\end{center}
\caption{Top panel: Phased radial velocity curve of all published radial velocity measurements of WASP-14. Bottom panel: Corresponding residuals after the radial velocity solution for the transiting hot Jupiter is removed. There is no significant linear acceleration detected in the data.} \label{rv}
\end{figure}

By combining the updated transit ephemeris and secondary eclipse times derived from the global phase-curve fits with the radial velocity measurements analyzed in \citet{knutson2014}, we can obtain new estimates of the orbital eccentricity and pericenter longitude: $e=0.0830^{+0.0029}_{-0.0030}$ and $\omega=252.67^{+0.70}_{-0.77}$~degrees. These values are consistent with those reported in \citet{knutson2014}. We use the updated values of orbital eccentricity and pericenter longitude in our final phase curve fits (Table~\ref{tab:values}). Results from our radial velocity fits are shown in Table~\ref{tab:rv} and Figure~\ref{rv}. These fits provide new estimates of the orbital period ($P_{b}$) and center of transit time ($T_{c,b}$), as well as the semi-amplitude of the planet's radial velocity ($K_{b}$), the radial velocity zero point offsets for data collected by each of the different spectrographs from which radial velocity measurements of the system were obtained ($\gamma_{1-4}$), and the slope ($\dot{\gamma}$) of the best-fit radial velocity acceleration. The radial velocity slope is consistent with zero, indicating no evidence for additional planets in the WASP-14 system.

From the RV fits we also arrive at updated values of the orbital semi-major axis and planet mass: $a=0.0371\pm 0.0011$~AU and $M_{p}=7.76\pm 0.47~M_{\mathrm{Jup}}$. Using the error-weighted best-fit values of $a/R_{*}$ and $R_{p}/R_{*}$ from both bandpasses, we obtain a new estimate of the planet's radius: $R_{p}=1.221\pm 0.041~R_{\mathrm{Jup}}$. A full list of updated planetary parameters is give in Table~\ref{tab:newvalues}.

\begin{table}[t!]
\centering
\begin{threeparttable}
\caption{Updated Planetary Parameters} \label{tab:newvalues}

\renewcommand{\arraystretch}{1.2}

\begin{tabular}{ l m{0.001cm} c  }
\hline\hline
 \multicolumn{1}{c}{Parameter}& & \multicolumn{1}{c}{Value} \\
  
 \hline 
\vspace{1mm}
$R_{p}/R_{*}$ & & $0.09419\pm 0.00043$ \\

$a/R_{*}$ & & $5.99\pm 0.09$\\
$i$~$(^{\circ})$ & & $84.63\pm 0.24$\\
$e$ & & $0.0830^{+0.0029}_{-0.0030}$ \\
$\omega$~$(^{\circ})$& & $252.67^{+0.70}_{-0.77}$ \\
$M_{p}$~$(M_{\mathrm{Jup}})$ & & $7.76\pm 0.47$ \\
$R_{p}$~$(R_{\mathrm{Jup}})$ & & $1.221\pm 0.041$\\
$\rho_{p}$~(g cm$^{-3}$) & & $5.29\pm 0.62$\\
$g_{p}$~(m s$^{-2}$) & & $129\pm 12$\\
$a$~(AU) & & $0.0371\pm 0.0011$ \\
\hline\hline

\end{tabular}
      \end{threeparttable}
\end{table}

\subsection{Phase curve fits}

\begin{figure*}[t*]
\begin{center}
\includegraphics[width=18cm]{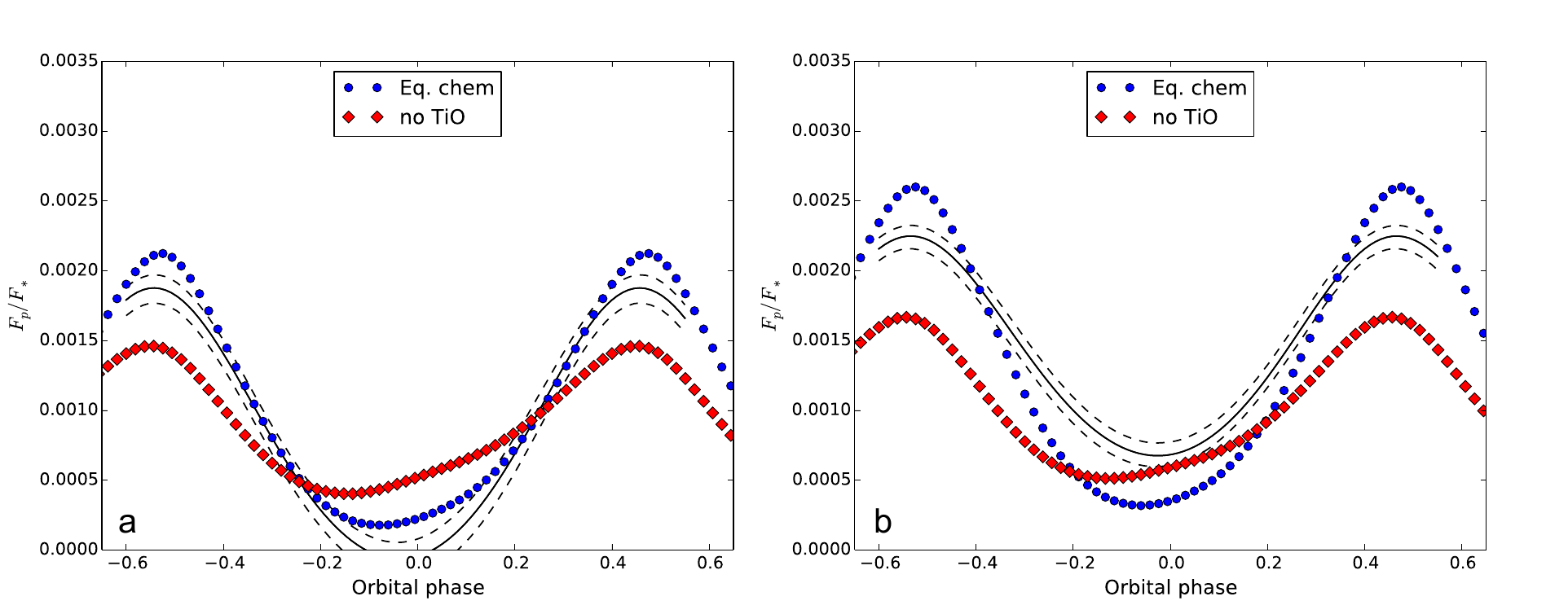}
\end{center}
\caption{(a) The best-fit 3.6~$\mu$m planet-star flux ratio phase curve and $1\sigma$ brightness bounds (solid and dotted lines). Predicted phase curves from the equilibrium chemistry and no TiO SPARC models are plotted with blue circles and red diamonds, respectively. (b) Same as (a), but at 4.5~$\mu$m.} \label{lightcurve}
\end{figure*}

In this section, we combine the results of our global fits with model-generated spectra and light curves to provide constraints on the atmospheric properties of the planet. To compute the relative planet-star flux ratio phase curve $F_{p}/F_{*}$, we subtract the secondary eclipse depth from the best-fit phase curve and divide by the remaining flux measured at the center of secondary eclipse, which represents the brightness of the star alone. For the secondary eclipse depths in each bandpass, we take the error-weighted average of the depths listed in Table~\ref{tab:values}. The resulting phase curves are shown in Figure~\ref{lightcurve} along with the corresponding $\pm 1\sigma$ brightness bounds.

\LongTables
 \begin{deluxetable}{lll}
     \tablecaption{Phase Curve Comparison}
    \tablehead{\colhead{Source} & \colhead{3.6~$\mu$m} & \colhead{4.5~$\mu$m}}
    \startdata
	\sidehead{\textit{Maximum flux ratio [$\%$]}}Measured & $0.1877^{+0.0094}_{-0.0108}$ & $0.2249^{+0.0077}_{-0.0090}$\\
Eq. chem. model & $0.2124$ & $0.2600$\\
No TiO model & $0.1461$ & $0.1667$\\
\\
\sidehead{\textit{Minimum flux ratio [$\%$]}}Measured & $< 0.0175$\tablenotemark{a} & $0.0675^{+0.0092}_{-0.0078}$\\
Eq. chem. model & $0.0178$ & $0.0318$\\
No TiO model & $0.0403$ & $0.0512$\\
\\
\sidehead{\textit{Phase curve amplitude [$\%$]}\tablenotemark{b}}Measured & $>0.1702$\tablenotemark{a} & $0.1574^{+0.0044}_{-0.0048}$\\
Eq. chem. model & $0.1946$ & $0.2282$\\
No TiO model &$0.1058$ & $0.1155$\\
\\
\sidehead{\textit{Maximum flux offset [h]}\tablenotemark{c}}Measured & $-1.43\pm 0.21$ & $-1.01\pm 0.21$ \\
Eq. chem. model & $+0.3$ & $+0.2$\\
No TiO model & $-0.9$ & $-1.0$\\
\\
\sidehead{\textit{Minimum flux offset [h]}\tablenotemark{c}}Measured & $-2.03^{+0.42}_{-0.39}$ & $-1.39^{+0.43}_{-0.40}$\\
Eq. chem. model &$-3.7$ & $-2.9$\\
No TiO model & $-6.2$ & $-6.2$\\
 \enddata\label{tab:comparison}
 	\tablenotetext{a}{Values based off of $2\sigma$ upper limit of the flux ratio minimum.}
	\tablenotetext{b}{Difference between maximum and minimum flux ratios.}
	\tablenotetext{c}{The maximum and minimum flux offsets are measured relative to the center of secondary eclipse time and center of transit time, respectively, and are derived from the phase curve fit parameters. Negative time offsets of the maximum or minimum flux indicate an eastward shift in the location of the hot or cold region in the planet's atmosphere.}
\end{deluxetable}

Some of the main quantitative characteristics of the phase curves are summarized in Table~\ref{tab:comparison}. Notably, the fluxes at both \textit{Spitzer} wavebands peak significantly before secondary eclipse, implying that the hottest regions are shifted eastward from the substellar point.  This behavior appears to be common on hot Jupiters and, in addition to WASP-14b, has also been clearly observed on HD 189733b \citep{knutson2007,knutson2009a,knutson2012}, HD 209458b \citep{zellem}, HAT-P-2b \citep{lewis}, WASP-43b \citep{stevenson}, WASP-12b \citep{cowan2012}, and Ups And b \citep{crossfield}.  

This behavior was predicted before the \textit{Spitzer} era \citep{showmanguillot,cooper} and has now
been reproduced in a wide variety of general circulation models \citep[GCMs, e.g.,][]{heng2011a,heng2011b,perna2,lewis2010,rm2010,rm2012,rm2013,dobbs,showman2009,showman2015}. In these models, the eastward hotspot displacement results from eastward 
advection due to a fast, eastward-flowing jet stream centered at the equator. 
Given such a jet, a significant hot spot offset occurs under conditions when the 
radiative timescale at the photosphere is comparable to timescales for air to 
advect horizontally over a planetary radius.  The eastward offsets in our 
observations thus provide evidence that WASP-14b exhibits an eastward 
equatorial jet stream.

It is interesting to compare our observed day-night flux differences with
those of other planets observed to date.  As mentioned previously, \citet{cowanagol2011} and \citet{perez-becker} inferred that, in general,
planets that receive higher stellar fluxes exhibit larger fractional day-night
temperature differences and less efficient day-night heat redistribution than 
planets that receive lower stellar fluxes. Averaging the phase curve amplitudes at 3.6 and 4.5~$\mu$m (listed in Table~\ref{tab:comparison}), we find that WASP-14b is roughly consistent with this trend,
with a day-night flux difference that is smaller than those of highly-irradiated planets like
WASP-12b and WASP-18b, but larger than those of more weakly-irradiated planets
like HD 189733b and HD 209458b.

\subsection{Brightness temperature}

We consider four types of atmosphere models. First, we use an interpolated PHOENIX spectrum for the host star WASP-14 \citep{husser} to calculate the brightness temperatures of WASP-14b in each band from the retrieved secondary eclipse depths. The best-fit brightness temperatures are $2405^{+29}_{-30}$~K at 3.6~$\mu$m and $2393^{+52}_{-50}$~K at 4.5~$\mu$m.  We find that the 3.6 and 4.5~$\mu$m eclipse depths are consistent with a single blackbody temperature and derive an effective temperature of $T_{\rm eff} = 2402\pm 35$~K from a simultaneous fit to both bandpasses. The predicted equilibrium temperature for WASP-14b assuming zero albedo is 2220~K if incident energy is re-radiated from the dayside only and 1870~K if the planet re-radiates the absorbed flux uniformly over its entire surface. We can also compare the computed brightness temperatures to the effective temperature of the dayside in the no-albedo, no-circulation limit assuming each region is a blackbody locally in equilibrium with the incident stellar flux: 2390~K \citep{cowanagol2011}. The high brightness temperatures from the blackbody fits therefore suggest that WASP-14b has a very hot dayside atmosphere and inefficient day-night recirculation.

\subsection{Dynamical models}

Next, we compare our phase curves and emission spectra to theoretical models generated from the three-dimensional Substellar and Planetary Atmospheric Radiation and Circulation (SPARC) model to investigate the global circulation of WASP-14b.  The SPARC model was specifically developed with the study of extrasolar planetary atmospheres in mind \citep{showman2009}. The SPARC model couples the MITgcm \citep{adcroft} with the non-gray radiative transfer model of \citet{marley} in order to self-consistently calculate the amount of heating/cooling at each grid point. In this way, the SPARC model does not require advective tuning parameters often employed in one-dimensional radiative transfer models or prescribed pressure-temperature profiles utilized in Newtonian cooling schemes employed in other circulation models.  Our use of an atmospheric model that fully couples radiative and dynamical processes is especially important for planets on eccentric orbits, such as WASP-14b, which experience time variable heating \citep{lewis2010,lewis2014,kataria}. 

\begin{figure}[t*]
\begin{center}
\includegraphics[width=9cm]{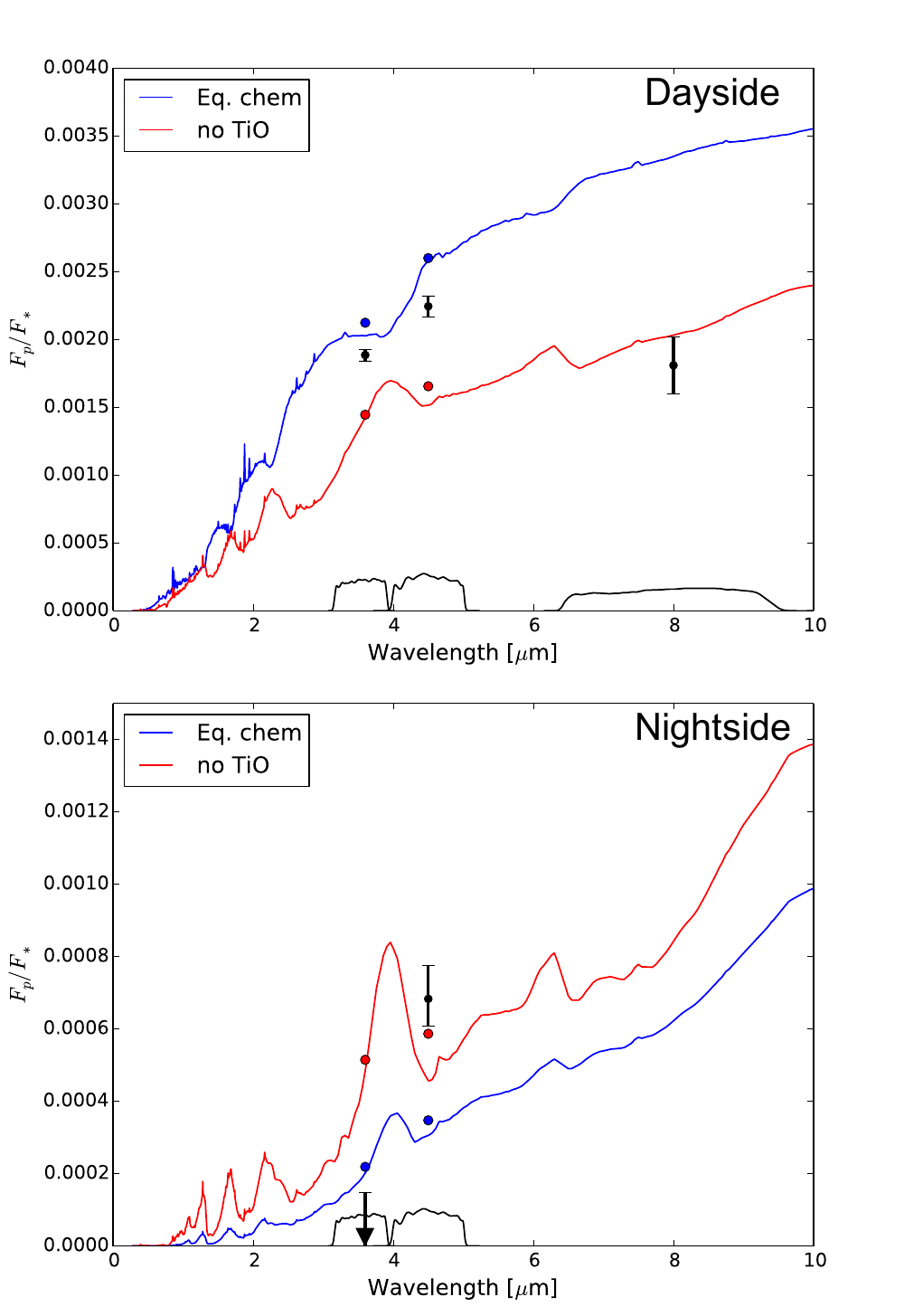}
\end{center}
\caption{Comparison of the error-weighted average broadband planet-star ratio measured at 3.6, 4.5, and 8.0~$\mu$m (filled black circles) with SPARC model emission spectra at the time of secondary eclipse and the time of transit, corresponding to the dayside and nightside of the planet. For the measured nightside 3.6~$\mu$m emission, the $2\sigma$ upper limit is shown. Solid lines indicate the predicted spectra for the equilibrium chemistry (blue) and no TiO models (red). Band-averaged fluxes are overplotted as filled points of the same color. The black lines at the bottom represent the photometric band transmission profiles, in arbitrary units. The measured dayside 3.6 and 4.5~$\mu$m planetary fluxes are bounded by the two models, suggesting a possible sub-solar abundance of TiO/VO; neither model reproduces the low measured 8.0~$\mu$m dayside emission, which is derived from a single secondary eclipse measurement previously published in \citet{blecic}. Meanwhile, the measured nightside planetary fluxes in both the 3.6 and 4.5~$\mu$m bands are highly discrepant from the model-predicted values, which may point toward an enhanced atmospheric C/O ratio.} \label{jonathan}
\end{figure}

\begin{figure}[t*]
\begin{center}
\includegraphics[width=9cm]{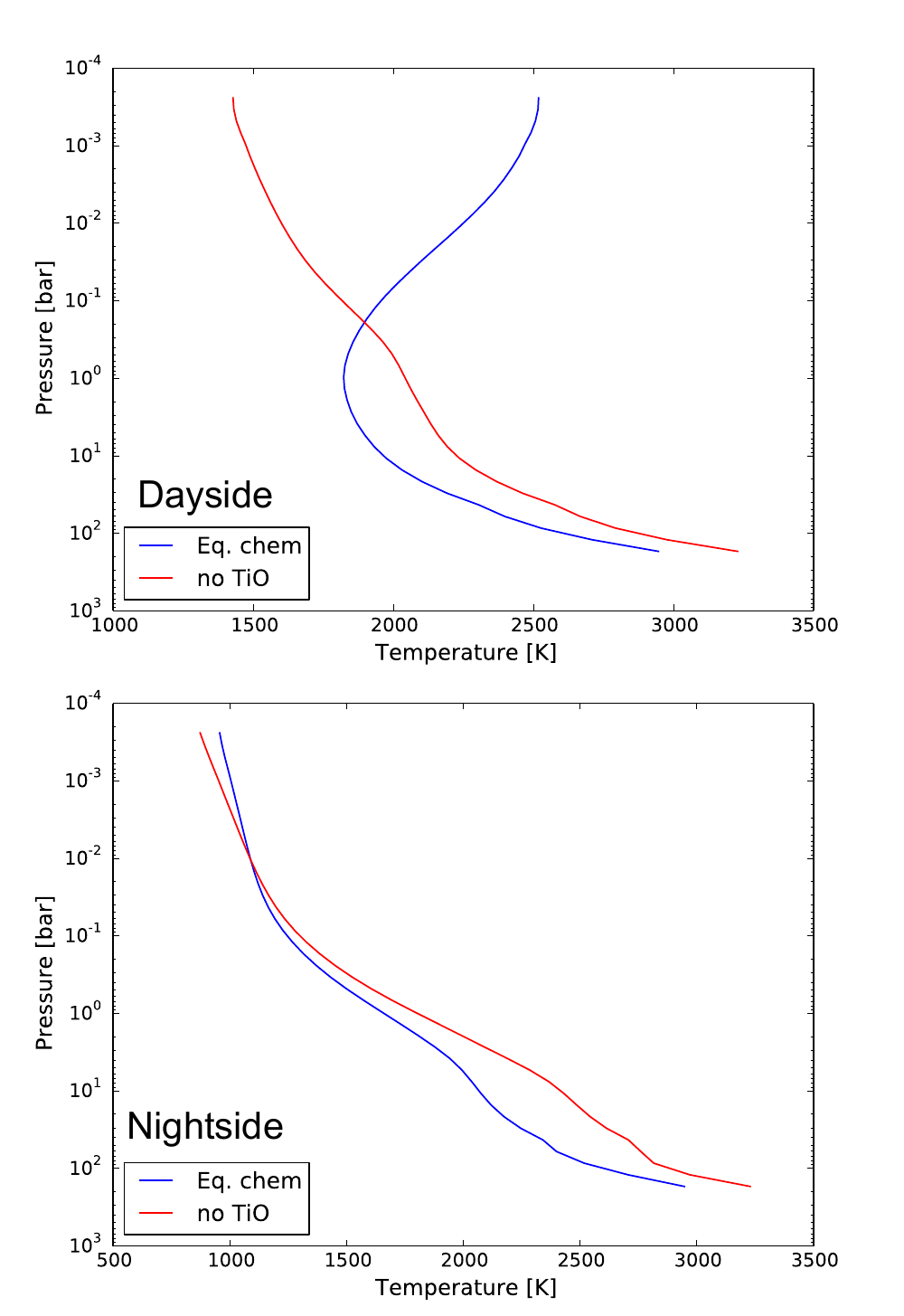}
\end{center}
\caption{Plot of the dayside and nightside temperature-pressure profiles computed by the SPARC models that correspond to the emission spectra in Figure~\ref{jonathan}.} \label{jonathantp}
\end{figure}

The SPARC models of WASP-14b presented here utilize the cubed-sphere grid \citep{adcroft} with 
a horizontal grid resolution of roughly $5.625^{\circ}$ in latitude and longitude (a so-called low-resolution C16 grid).  
In the vertical direction, the models span pressures ranging from 200~bar to 0.2~mbar with 39 layers evenly 
spaced in $\log{p}$ and a top layer that extends to zero pressure.  Here we consider atmospheric compositions 
in thermochemical equilibrium both with and without the incorporation of the strong optical absorbers TiO and VO (hereafter, ``equilibrium chemistry'' and ``no TiO'', respectively).  The presence of TiO/VO in our models allows for the development of a dayside inversion layer \citep{fortney}. All models adopt solar elemental ratios of heavy elements.

We assume that WASP-14b is in a pseudo-synchronous rotation state \citep[P$_{rot}\sim 2.14$ days;][]{hut}.  As in the case of GJ~436b,
WASP-14b's relatively low eccentricity means that assuming pseudo-synchronous rotation instead of synchronous rotation is not likely to strongly affect 
the global circulation patterns that develop \citep{lewis2010}. 

Figure~\ref{lightcurve} compares the best-fit phase curve derived from the \textit{Spitzer} data in each bandpass with the band-averaged light curves generated from the SPARC model using the methods of \citet{fortney2006}. Model dayside and nightside spectra at the center of secondary eclipse and center of transit times are shown in Figure~\ref{jonathan} along with the corresponding measured flux ratios. The corresponding dayside and nightside temperature-pressure profiles generated by the models are shown in Figure~\ref{jonathantp}. For the dayside planetary emission, we combine the 3.6 and 4.5~$\mu$m flux ratios reported in the present work and previously in \citet{blecic} to arrive at the error-weighted average values: $0.1886\%\pm 0.0043\%$ at 3.6~$\mu$m and  $0.2245\%\pm 0.0078\%$ at 4.5~$\mu$m. We also include the measured 8.0~$\mu$m flux ratio ($0.181\% \pm 0.021\%$) from \citet{blecic}. Table~\ref{tab:comparison} compares the model-predicted maximum and minimum flux ratios and time offsets with the values derived from the data. 

In both the 3.6 and 4.5~$\mu$m bandpasses, the measured dayside planetary emission lies between the equilibrium chemistry and no TiO models. A possible explanation is that the dayside atmosphere of WASP-14b may contain a sub-solar abundance of TiO/VO (possibly due to cold trapping), which would result in a weaker temperature inversion than the one predicted by the solar abundance equilibrium chemistry model \citep[e.g.,][]{fortney,spiegel,madhusudhan2011,parmentier}. Meanwhile, the measured 8.0~$\mu$m dayside brightness is not reproduced by either of the models.

We report offsets in the time of maximum and minimum flux relative to the center of transit and center of secondary eclipse times, respectively, as derived from our phase curve analysis and the models (Table~\ref{tab:comparison}). Negative time offsets of the maximum or minimum flux indicate an eastward shift in the location of the hot or cold region in the planet's atmosphere. The no TiO model greatly overestimates the magnitude of the minimum flux offsets while giving good agreement with the maximum flux offsets at both wavelengths. The equilibrium chemistry model yields a less severe overestimation of the minimum flux offset, while predicting a maximum flux offset that is opposite the one measured from the data. We find that overall the equilibrium chemistry model yields the closer match. The no TiO model predicts a significant asymmetry in the phase curve around the time of minimum flux, which is not seen in the \textit{Spitzer} data.

The most notable discrepancy between the best-fit phase curves and the SPARC model results is in the nightside planetary flux. Both the equilibrium chemistry and the no TiO model overestimate the planet's nightside brightness at 3.6~$\mu$m, while underestimating the brightness at 4.5~$\mu$m. The planet's very low 3.6~$\mu$m nightside emission indicates a higher atmospheric opacity at that wavelength than is predicted by the models. Meanwhile, the higher-than-predicted 4.5~$\mu$m nightside emission points toward a slight reduction in the atmospheric opacity at that wavelength relative to the models. One possible explanation for both of these trends is an increased C/O ratio. \citet{moses} demonstrate that increasing the C/O ratio above equilibrium values leads to an excess of CH$_{4}$ and a depletion of CO. This enhances the opacity in the 3.6~$\mu$m bandpass where CH$_{4}$ has strong vibrational bands, while reducing the opacity in the 4.5~$\mu$m bandpass where CO has strong vibrational bands.

An enhanced C/O ratio can also be invoked to explain the higher amplitude of the 3.6~$\mu$m phase curve as compared with the 4.5~$\mu$m phase curve. In this scenario the 3.6~$\mu$m bandpass would probe lower pressure regions that have strong day/night contrasts due to a combination of short radiative timescales and the possible presence of a dayside temperature inversion. A relative depletion of CO would shift the 4.5~$\mu$m photosphere to higher pressures, where longer radiative timescales facilitate more efficient day-night heat transport, resulting in a reduced phase curve amplitude. We note, however, that recent spectroscopic surveys of M dwarfs suggest that the occurrence of high C/O ratios in stellar atmospheres appears to be low \citep[see][and references therein]{gaidos}. At the same time, models of disk chemistry predict that the C/O ratio of the solids and gas in the disk may vary as a function of disk radius \citep[e.g.,][]{oberg,madhusudhan}. In particular, different snowlines of oxygen- and carbon-rich ices are expected to result in systematic variations in the C/O ratio of gaseous material across the protoplanetary disk. Gas giants that derive most of their atmosphere from the gas disk outside of the water snowline but inside of the carbon dioxide snowline may therefore end up with high atmospheric C/O ratios. We also note that \citet{madhusudhan2011} showed that planets with high C/O ratios should naturally have less TiO and correspondingly weaker temperature inversion, consistent with our previous discussion of WASP-14b's dayside emission. In addition, \citet{blecic} show that atmospheric models with high C/O ratios often predict reduced emission at longer wavelengths ($6-8$~$\mu$m) and may explain the low 8.0~$\mu$m planetary flux ratio (see Figure~\ref{jonathan}).

An alternative explanation for the planet's low 3.6~$\mu$m nightside emission is the presence of high-altitude silicate clouds on the cold nightside in or above the photosphere. 
The formation of an equilibrium silicate cloud is possible at the
photospheric pressures ($\sim$100~mbar) probed on the nightside of
WASP-14b by these observations \citep[e.g.][]{visscher2010}.  However, it
is expected that the presence of a thick cloud would suppress both the 3.6
and 4.5~$\mu$m nightside fluxes from the planet, and the resulting planetary emission spectrum would resemble a blackbody with an effective temperature corresponding to that of the cloud tops. While the low 3.6~$\mu$m nightside brightness temperature of $1079$~K ($2\sigma$ upper limit) is consistent with a cold emitting layer, the two nightside band fluxes together are not consistent with a single blackbody spectrum. 

In this scenario, the high 4.5~$\mu$m nightside emission (corresponding to a brightness temperature of $1380^{+70}_{-60}$~K) could be explained by introducing an emitting layer of CO in a warmer thermosphere that is situated above the cold cloud tops \citep{koskinen2013}. However, it is unclear how a high-altitude temperature inversion might arise on the nightside, as there is no incident stellar irradiation and the efficiency of recirculation from the dayside should be relatively weak at low pressures. Therefore,  an enhanced C/O ratio in WASP-14b's atmosphere provides a more straightforward explanation for the observed
differences between WASP-14b's 3.6 and 4.5~$\mu$m phase curves.  Further
observations of WASP-14b, possibly with {\it Hubble Space Telescope}'s Wide Field Camera 3, would provide
important additional information to constrain WASP-14b's composition and
either support or refute the high atmospheric C/O ratio scenario posited
here for WASP-14b.  We note that \citet{lewis2014} also suggested an
enhanced C/O ratio to explain the differences seen between the 3.6 and
4.5~$\mu$m phase-curve observations of HAT-P-2b.  Additional phase-curve
observations of hot Jupiters may reveal similar trends and point to a
fundamental piece of physics currently missing from exoplanet atmospheric
models and/or planet formation theories. The SPARC model utilized in this work  does not readily accommodate non-solar C/O ratios. The exploration of the effects of different C/O ratios on the atmospheric dynamics of hot Jupiters will be the topic of future work.

We also compare the \textit{Spitzer} dayside and nightside emission with 1D models generated using the methods of \citet{burrows} with varying recirculation. These models assume local thermodynamic equilibrium, solar composition, and a plane-parallel atmosphere. A heat sink is included at depth (between 0.003 and 0.6 bars) to redistribute heat from the dayside to the nightside. These models also incorporate a generalized gray absorber at low pressures (0 to 0.03 bars), which is parametrized with an absorption coefficient, $\kappa_{e}$, with units of cm$^{2}$~g$^{-1}$.  This absorber enhances the opacity of the planet's atmosphere at optical wavelengths, raising the local atmospheric temperature and producing a high-altitude temperature inversion. A second dimensionless parameter $P_{n}$ is used to specify the efficiency of energy redistribution, with $P_{n}=0.5$ indicating complete redistribution and $P_{n}=0$ signifying redistribution on the dayside only.

\begin{figure}[t*]
\begin{center}
\includegraphics[width=9cm]{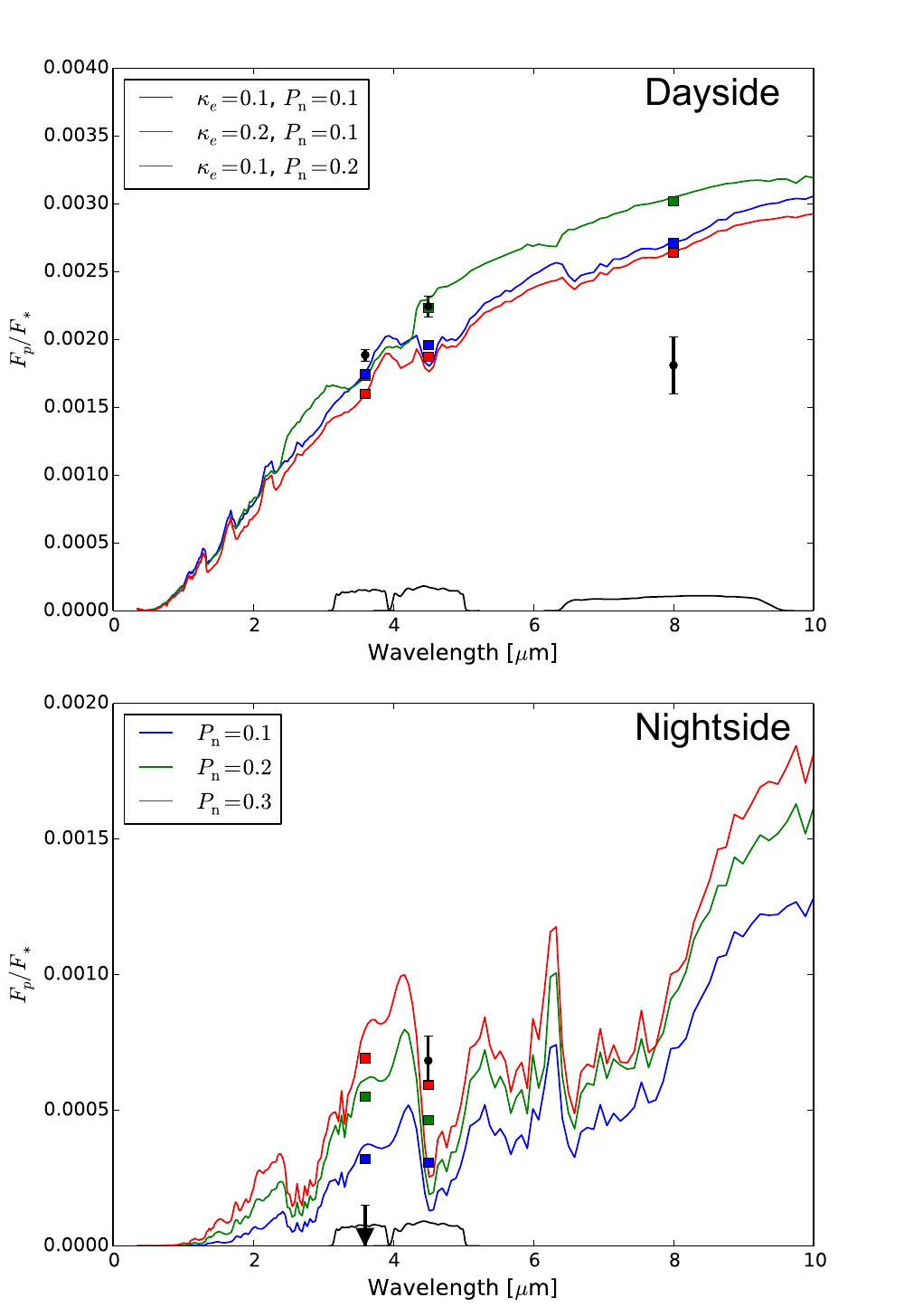}
\end{center}
\caption{Comparison of the error-weighted average broadband planet-star ratio measured at 3.6, 4.5, and 8.0~$\mu$m (filled black circles) with one-dimensional atmosphere model spectra following \citet{burrows} at the time of secondary eclipse and the time of transit, corresponding to the dayside and nightside of the planet. For the measured nightside 3.6~$\mu$m emission, the $2\sigma$ upper limit is shown. Solid colored lines indicate the predicted spectra for various choices of the parameters $\kappa_{e}$ and $P_{n}$, which represent the abundance of a generalized gray absorber in the upper atmosphere and the efficiency of energy redistribution, respectively. Corresponding band-averaged points are overplotted in the same color. The black lines at the bottom represent the photometric band transmission profiles, in arbitrary units. The measured dayside 3.6 and 4.5~$\mu$m planetary fluxes are most consistent with the model with $\kappa_{e}=0.2$ and $P_{n}=0.1$, indicating that WASP-14b has poor day-night recirculation and a moderate thermal inversion in the dayside atmosphere; none of the models reproduces the low measured 8.0~$\mu$m dayside emission. On the nightside, the measured nightside planetary fluxes in both the 3.6 and 4.5~$\mu$m bands are not well-described by any of the models, which may point toward an enhanced atmospheric C/O ratio.} \label{burrows}
\end{figure}

\begin{figure}[t*]
\begin{center}
\includegraphics[width=9cm]{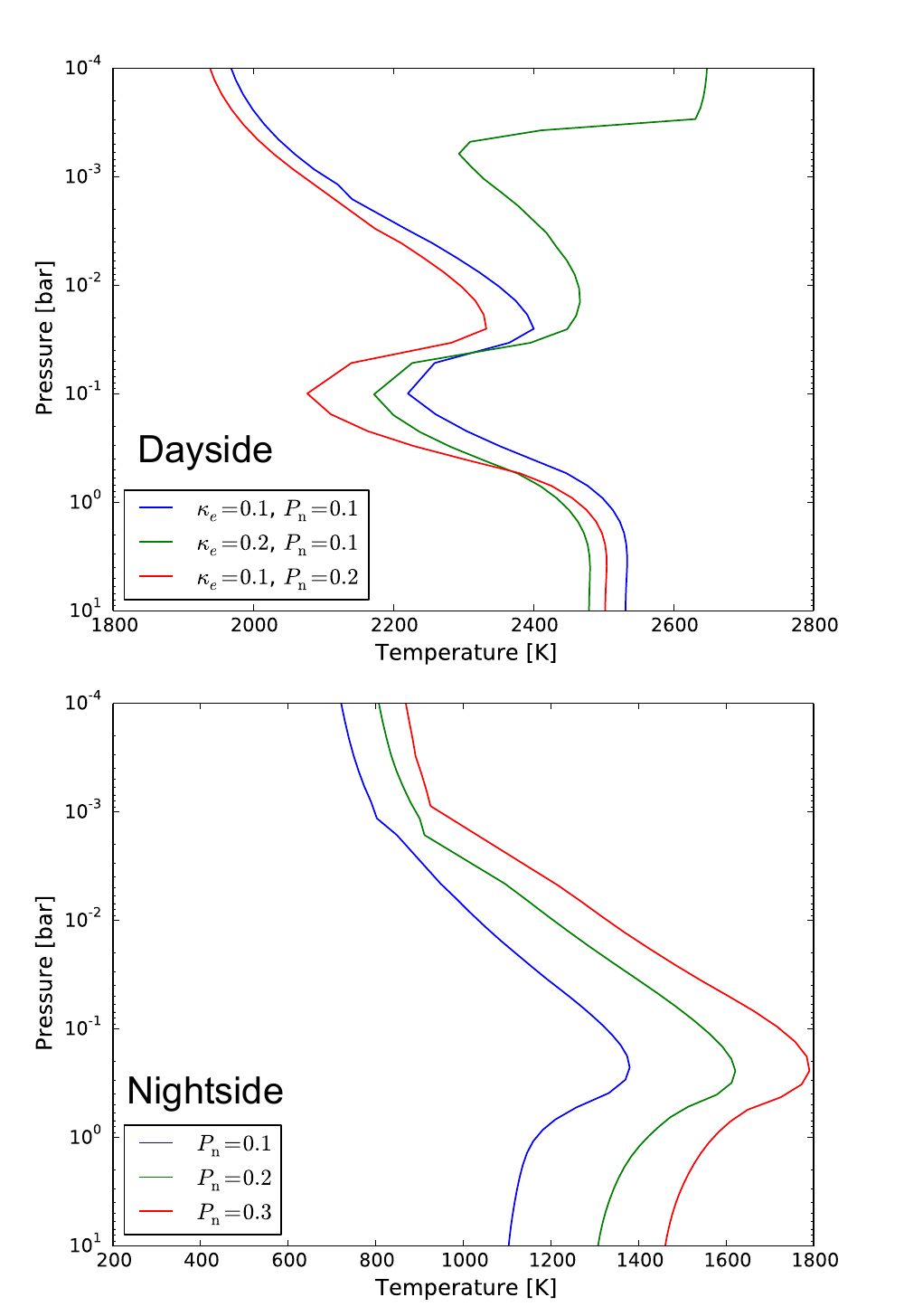}
\end{center}
\caption{Plot of the dayside and nightside temperature-pressure profiles computed by the one-dimensional atmosphere model used to generate the spectra shown in Figure~\ref{burrows}.} \label{burrowstp}
\end{figure}

In Figure~\ref{burrows}, we show the dayside and nightside spectra for various values of $\kappa_{e}$ and $P_{n}$ and compare them with the measured relative planetary brightnesses in the two \textit{Spitzer} bands. Figure~\ref{burrowstp} shows the corresponding dayside and nightside temperature-pressure profiles generated by the model. In these models the presence of the gray absorber does not affect day-night recirculation, so the nightside spectra depend only on the value of $P_{n}$. We define the measured dayside brightness in each bandpass to be the measured secondary eclipse depth, while for the the nightside data points, we used the value of the best-fit relative planetary phase curve $F_{p}/F_{*}$ calculated at the mid-transit time in each bandpass. Looking at the dayside spectra, we see that the $\kappa_{e}=0.2$ and $P_{n}=0.1$ model spectrum comes closest to matching the measured data points at 3.6 and 4.5~$\mu$m. This suggests that WASP-14b has poor day-night recirculation and a moderate thermal inversion in the dayside atmosphere.

On the nightside, as in the SPARC models, none of the models reproduce the low measured 3.6~$\mu$m brightness, while the measured 4.5~$\mu$m flux ratio is most consistent with the $P_{n}=0.3$ model. As discussed earlier in the context of the SPARC models, this discrepancy is consistent with the hypothesis that WASP-14b might have an enhanced C/O ratio. We also note that large deviations from solar-like equilibrium chemical composition could yield a larger disparity in the pressures probed on the dayside and nightside than is accounted for in our models. Specifically, one could be probing a deeper than expected pressure level on the nightside, where circulation is efficient (high $P_{n}$), while probing a higher-up (lower pressure) level on the dayside, where the circulation is less efficient (low $P_{n}$). This scenario is especially probable when there is a dayside temperature inversion, as our data suggest.

\subsection{Albedo}

\begin{figure}[t*]
\begin{center}
\includegraphics[width=9cm]{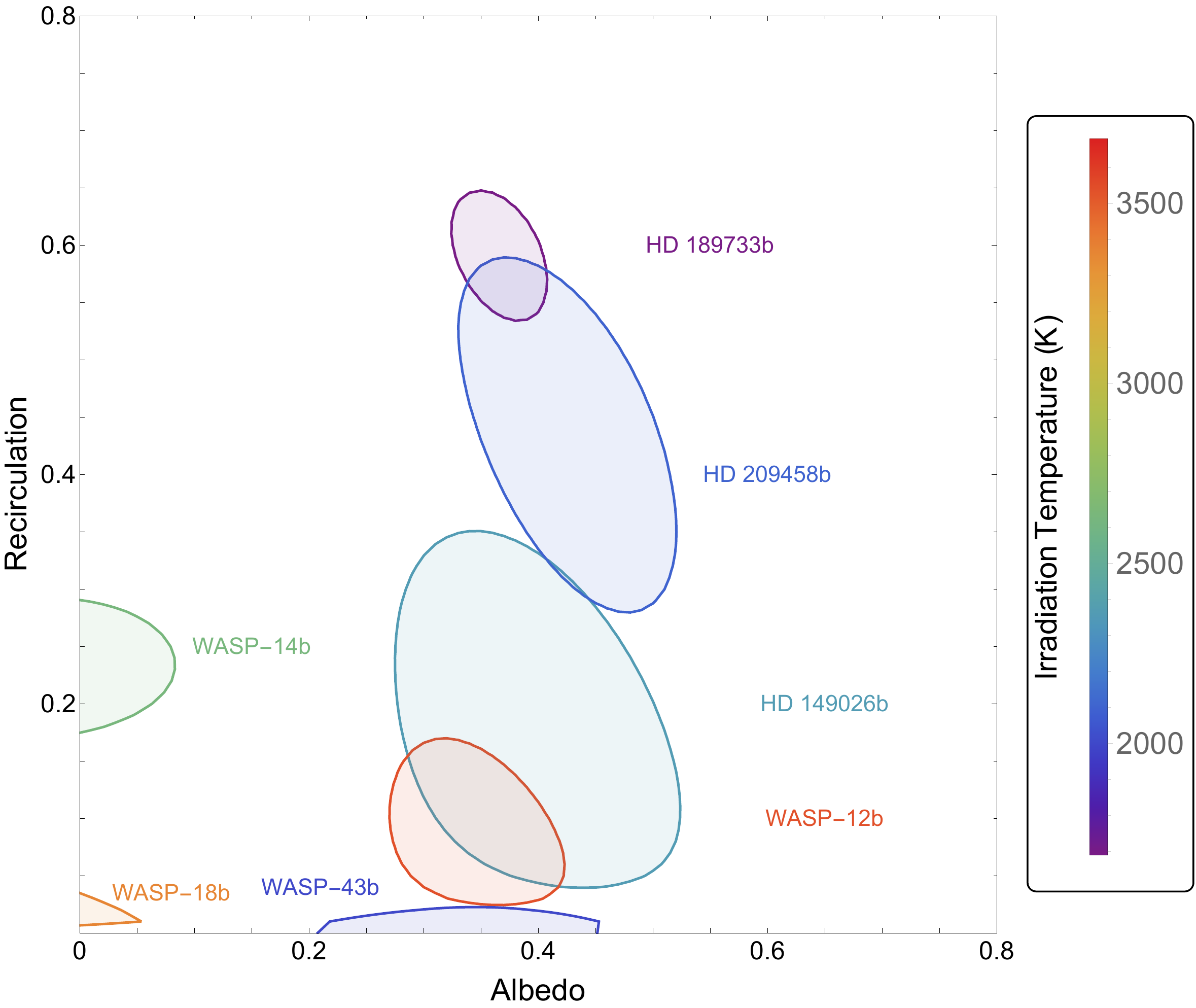}
\end{center}
\caption{Comparison of composite $1\sigma$ Bond albedo and recirculation efficiency confidence regions for planets with thermal phase curve observations, as calculated using the methods of \citet{schwartz}. The color of the bounding curves indicate irradiation temperature. WASP-14b has intermediate day-night heat transport, like other hot Jupiters of similar temperature. However, it has a very low Bond albedo, like WASP-18b. Both of these planets are significantly more massive than the other planets in the sample, suggesting that they may be emitting detectable heat from the deep atmosphere or interior processes.} \label{planets}
\end{figure}

Finally, we compare the albedo and recirculation derived from our best-fit phase curves to other hot Jupiters for which full-orbit thermal measurements have been obtained. Following the methods described in \citet{schwartz}, we use the measured eclipse depths and phase curve amplitudes to calculate the error-weighted dayside and nightside brightness temperatures, with corrections for the contamination due to reflected light: $T_{d}=2312\pm 35$~K and $T_{n}=1299\pm 77$~K. Although these methods were developed for planets on circular orbits, the low eccentricity of WASP-14b places it in a regime for which the model is still a reasonable approximation. From these temperatures, we derive a Bond albedo of $A_{B} = 0$ ($<$0.08 at $1\sigma$) and a day-night heat transport efficiency of $\epsilon=0.23\pm 0.06$, where the latter is defined such that $\epsilon=0$ means no heat recirculation to the nightside, and $\epsilon=1$ indicates complete redistribution. Figure~\ref{planets} shows the location of WASP-14b in albedo-recirculation space along with six other exoplanets with measured thermal phase curves. WASP-14b's low day-night heat transport efficiency and high irradiation temperature are consistent with the general observed trend that highly-irradiated hot Jupiters have poor heat recirculation \citep{cowanagol2011,perez-becker}.

Both WASP-14b and WASP-18b have much lower estimated albedos than those of other Jovian mass planets with thermal phase curve measurements; they are also both significantly more massive ($7-10$ vs. $\sim 1~M_{\mathrm{Jup}}$). The low estimated albedos of these massive planets, implied by their high thermal emission, may instead be indicative of the detection of some amount of thermal radiation from the planet's interior in addition to the re-radiated stellar flux.  A different cooling and/or migration history for these massive planets could result in an added contribution to the planetary emission from the deep atmosphere; the age of the WASP-14 system is relatively young \citep[$<$1~Gyr;][]{joshi}, so this additional internal flux may be due to significant residual heat of formation. In addition, the higher internal fluxes of these higher-mass planets may support stronger magnetic fields than on smaller planets, which may lead to stronger Ohmic dissipation in their atmospheres \citep[e.g.,][]{christensen,batygin2013}, thereby providing a source of additional emission on the dayside.

\section{Conclusions}
In this paper we present the first phase curve observations of the eccentric hot Jupiter WASP-14b in the \textit{Spitzer} 3.6 and 4.5~$\mu$m bandpasses. We compare two different techniques --- pixel mapping and pixel-level decorrelation --- for correcting the intrapixel sensitivity effect and find that the pixel mapping method yields lower residual scatter from the best-fit solution. This is likely due to the relatively large range of star positions on the pixel throughout the full-orbit observations that may make the intrapixel sensitivity effect less amenable to modeling through pixel-level decorrelation. We obtain best-fit secondary eclipse depths of  $0.1882\%\pm 0.0048\%$ and $0.2247\%\pm 0.0086\%$ at 3.6 and 4.5~$\mu$m, respectively, which are consistent with a single blackbody brightness temperature of $2402\pm 35$~K. These depths are in good agreement (within $1\sigma$) with the ones reported by \citet{blecic} in their \textit{Spitzer} secondary eclipse analysis. Combining the results of our global phase curve fits with previous radial velocity measurements, we derive updated values for orbital inclination, orbital eccentricity, longitude of pericenter, orbital semi-major axis, planet radius, and planet mass. We also combine our measured transit times with previously-published transit times to arrive at a more precise estimate of WASP-14b's orbital period: $P=2.24376507\pm 0.00000046$~days.

Comparison of the measured dayside planetary emission with spectra generated from a one-dimensional radiative transfer model \citep{burrows} suggests relatively inefficient day-night heat recirculation and a moderate dayside temperature inversion. The relatively high dayside blackbody temperature provides additional support for the idea that the day-night circulation is inefficient. The flux maxima precede the secondary eclipses, consistent with other hot Jupiter lightcurves and with predictions of general circulation models (GCMs), suggesting the possibility of equatorial superrotation on WASP-14b. We also utilize a three-dimensional GCM \citep{showman2009} to generate theoretical light curves for an atmosphere in thermochemical equilibrium both with and without a dayside temperature inversion. We find that the measured amplitude and location of minimum/maximum flux in both bandpasses are more consistent with predictions from the model light curve with a dayside thermal inversion. Meanwhile, the measured nightside planetary emission at 3.6 and 4.5~$\mu$m is not adequately described by either the one-dimensional or the three-dimensional models. In particular, the very low 3.6~$\mu$m nightside planetary flux indicates a significantly higher atmospheric opacity at that wavelength than is predicted by the models, which may point toward an enhanced C/O ratio. In the context of other planets with full-orbit thermal measurements, we find that WASP -14b fits the general trend that highly-irradiated hot Jupiters have poor heat recirculation, while the derived Bond albedo is very small ($<$0.08 at $1\sigma$) and the planet's large mass might indicate that WASP-14b is emitting residual heat from its formation.

The question of whether WASP-14b has an enhanced C/O ratio can be further addressed by obtaining measurements of the host star's C/O ratio. We note that WASP-14 has a near-solar metallicity \citep[$\lbrack$Fe/H$\rbrack=-0.18\pm 0.08$;][]{torres}, so from a statistical standpoint, it is not expected to have a high C/O ratio \citep{teske}. Future work will explore how the assumed C/O ratio in the atmospheres of hot-Jupiters like WASP-14b affects global circulation patterns and
day/night temperature contrasts. The discrepancy between the model-predicted and measured nightside planetary flux underlines the need for further exploration of the available parameter space in both 1D and GCM models of hot Jupiter atmospheres. Assessing the effects of non-solar chemistry, and specifically different C/O ratios, will give us new insight into the interplay between various atmospheric properties and the resultant planetary emission. These studies promise to greatly enhance the ability of numerical models to explain features in the growing body of exoplanet phase curves. 

Obtaining more infrared phase curves of massive planets and calculating their Bond albedo will allow us to determine whether low Bond albedo is strongly correlated with high planet mass, as is the case for WASP-14b and WASP-18b, and further consider the contribution of residual heat from formation in the overall emission of massive hot Jupiters. A better understanding of the role of residual heat from formation in the flux of massive planets is also important for directly-imaged planets, as cooling curves are used to estimate the planetary mass based on the age of the system and the measured luminosity \citep[e.g.,][]{fortney2}. If we can determine whether current cooling models can reproduce the observed emission from WASP-14b, it would serve as an independent confirmation of the evolutionary models used in studies of directly imaged planets.

\section*{Acknowledgments}
This work is based on observations made with the \textit{Spitzer
Space Telescope}, which is operated by the Jet Propulsion Laboratory,
California Institute of Technology under a contract
with NASA. Support for this work was provided by NASA through an award issued by JPL/Caltech. The authors wish to thank J.~I. Moses, J.~K. Teske, and T.~T. Koskinen for many useful discussions during the preparation of this manuscript. The authors also thank an anonymous reviewer for constructive comments that helped to improve the manuscript.

\end{document}